\documentclass[10pt]{iopart}
\lccode`\3`\3
\lccode`\/`\/
\usepackage[
    bottom=1.1cm
]{geometry}
\usepackage[british]{babel}
\usepackage{csquotes}

\expandafter\let\csname equation*\endcsname\relax
\expandafter\let\csname endequation*\endcsname\relax
\usepackage{amsmath}
\usepackage{amssymb}
\usepackage{dblfloatfix}
\usepackage[per-mode = symbol]{siunitx}
\usepackage{enumitem}
\usepackage{graphicx}
\usepackage{xcolor}
\usepackage{siunitx}
\usepackage[backend=biber, eprint=false, urldate=long, dateabbrev=false, style=ieee]{biblatex}
\AtEveryBibitem{%
  \ifentrytype{online}{%
  }{%
    \clearfield{howpublished}%
    \clearfield{url}%
    \clearfield{urldate}%
    \clearfield{issn}%
  }%
}
\usepackage[
    pdftitle={Transient Finite Element Simulation of Accelerator Magnets Using Thermal Thin Shell Approximation},
    pdfauthor={Erik Schnaubelt},
    pdfkeywords={quench simulation, finite element method, magneto-thermal analysis, thin-shell approximation, open-source},
    colorlinks,
    citecolor=blue,
    urlcolor=black,
    linkcolor=blue
]{hyperref}
\usepackage{droidsansmono}
\usepackage{subcaption}
\DeclareCaptionFormat{iop-caption}{\raggedright \footnotesize \textbf{#1#2}#3}
\DeclareCaptionFormat{iop-subcaption}{\centering\footnotesize #1#2#3}
\definecolor{BackgroundColor}{RGB}{250, 250, 250}
\definecolor{YamlKeyColor}{RGB}{75, 105, 198}
\definecolor{YamlValueColor}{RGB}{68,140,39}
\definecolor{YamlCommentColor}{RGB}{150, 150, 150}
\definecolor{YamlColonColor}{RGB}{119, 119, 119}
\usepackage{listings}
\usepackage[framemethod=TikZ]{mdframed}
\newmdenv[
    outerlinewidth=0.3,
    outerlinecolor=YamlCommentColor,
    middlelinewidth=0,
    backgroundcolor=BackgroundColor,
    roundcorner=3pt,
    innerbottommargin=0pt,
    innertopmargin=0pt,
    innerleftmargin=7pt,
    innerrightmargin=7pt,
    skipbelow=-0pt
]{yaml_box}
\newcommand\YAMLcolonstyle{\color{YamlColonColor}\droidsansmono\scriptsize\mdseries}
\newcommand\YAMLkeystyle{\color{YamlKeyColor}\droidsansmono\scriptsize\bfseries}
\newcommand\YAMLvaluestyle{\color{YamlValueColor}\droidsansmono\scriptsize\mdseries}
\makeatletter
\newcommand\language@yaml{yaml}
\expandafter\expandafter\expandafter\lstdefinelanguage
\expandafter{\language@yaml}
{
  keywords={true,false,null,y,n},
  keywordstyle=\YAMLkeystyle,
  basicstyle=\YAMLkeystyle,
  sensitive=false,
  comment=[l]{\#},
  morecomment=[s]{/*}{*/},
  commentstyle=\color{YamlCommentColor},
  stringstyle=\YAMLvaluestyle,
  moredelim=[l][\color{orange}]{\&},
  moredelim=[l][\color{magenta}]{*},
  moredelim=**[il][\YAMLcolonstyle{:}\YAMLvaluestyle]{:},
  moredelim=**[il][\YAMLcolonstyle{|}\YAMLvaluestyle]{|},
  morestring=[b]',
  morestring=[b]",
  literate =    {---}{{\ProcessThreeDashes}}3
                {>}{{\textcolor{red}\textgreater}}1     
                {|}{{\textcolor{red}\textbar}}1 
}
\lst@AddToHook{EveryLine}{\ifx\lst@language\language@yaml\YAMLkeystyle\fi}
\makeatother
\newcommand\ProcessThreeDashes{\llap{\color{cyan}\mdseries-{-}-}}
\usepackage{tudacolors}
\usepackage{pgfplots}
\usepackage{pgfplotstable}
\pgfplotsset{compat=1.18}
\usepgfplotslibrary{colormaps} 
\usetikzlibrary{shapes,intersections,positioning,calc,fit,patterns,patterns.meta,decorations.markings,decorations.text,arrows,arrows.meta,spy,calc}
\tikzset{>=latex}

\usepackage{calc}

\makeatletter
\tikzset{
  nomorepostactions/.code={\let\tikz@postactions=\pgfutil@empty},
  mymark/.style 2 args={decoration={markings,
    mark= between positions 0 and 1 step 0.04 with{%
        \tikzset{#2,every mark}\tikz@options
        \pgftransformresetnontranslations\pgfuseplotmark{#1}%
      },  
    },
    postaction={decorate},
    /pgfplots/legend image post style={
        mark=#1,mark options={#2},every path/.append style={nomorepostactions}
    },
  },
}
\makeatother

\pgfplotsset{
    colormap/plasma/.style={%
        /pgfplots/colormap={plasma}{%
          rgb=(0.050383, 0.029803, 0.527975)
          rgb=(0.186213, 0.018803, 0.587228)
          rgb=(0.287076, 0.010855, 0.627295)
          rgb=(0.381047, 0.001814, 0.653068)
          rgb=(0.471457, 0.005678, 0.659897)
          rgb=(0.557243, 0.047331, 0.643443)
          rgb=(0.636008, 0.112092, 0.605205)
          rgb=(0.706178, 0.178437, 0.553657)
          rgb=(0.768090, 0.244817, 0.498465)
          rgb=(0.823132, 0.311261, 0.444806)
          rgb=(0.872303, 0.378774, 0.393355)
          rgb=(0.915471, 0.448807, 0.342890)
          rgb=(0.951344, 0.522850, 0.292275)
          rgb=(0.977856, 0.602051, 0.241387)
          rgb=(0.992541, 0.687030, 0.192170)
          rgb=(0.992505, 0.777967, 0.152855)
          rgb=(0.974443, 0.874622, 0.144061)
          rgb=(0.940015, 0.975158, 0.131326)
    },
  },
}

\pgfplotsset{
    colormap/plasmaconstant/.style={%
        /pgfplots/colormap={plasmaconstant}{%
          rgb=(0.768090, 0.244817, 0.498465)
          rgb=(0.823132, 0.311261, 0.444806)
    },
  },
}

\newcommand{\equal}{=}

\begin{document}

\title{Transient Finite Element Simulation of Accelerator Magnets Using Thermal Thin Shell Approximation}

\author{Erik Schnaubelt\textsuperscript{1, 2}, Andrea Vitrano\textsuperscript{1}, Mariusz Wozniak\textsuperscript{1}, Emmanuele Ravaioli\textsuperscript{1}, Arjan Verweij\textsuperscript{1}, and Sebastian Schöps\textsuperscript{2}}

\address{\textsuperscript{1} CERN, Switzerland}
\address{\textsuperscript{2} Technical University of Darmstadt, Germany}
\ead{erik.schnaubelt@cern.ch}
\vspace{10pt}
\begin{indented}
\item[]\today
\end{indented}

\begin{abstract}
Thermal transient responses of superconducting magnets can be simulated using the finite element (FE) method. Some accelerator magnets use cables whose electric insulation is significantly thinner than the bare electric conductor. The FE discretisation of such geometries with high-quality meshes leads to many degrees of freedom. This increases the computational time, particularly since non-linear material properties are involved. In this work, we propose to use a thermal thin-shell approximation (TSA) to improve the computational efficiency when solving the heat diffusion equation in two dimensions. We apply the method to compute the thermal transient response of superconducting accelerator magnets used for CERN's Large Hadron Collider (LHC) and High-Luminosity LHC. The TSA collapses thin electrical insulation layers into lines while accurately representing the thermal gradient across the insulation's thickness. It allows considering cryogenic cooling via a temperature-dependent heat transfer coefficient and multi-layered quench heater (QH) regions. The TSA is implemented in the multipole module of the open-source Finite Element Quench Simulator (FiQuS), which can generate the multipole magnet models programmatically from input text files. The modifications implemented in FiQuS when constructing the geometry and mesh for the TSA model are discussed. First, the TSA approach is verified by comparison to classical FE simulations with meshed surface insulation regions for a simple block of four superconducting cables and a detailed model of the single-aperture Nb$_{3}$Sn dipole MBH. The results show that the TSA approach reduces the computational time significantly while preserving the accuracy of the solution. Second, the QH delay, i.e., the time between activation of the QH and the point in time where the transport current density exceeds the homogenized critical current density in one of the superconducting half turns, computed with the TSA method is compared to measurements for the MBH magnet. To this end, the thermal transient simulation is coupled to a magnetostatic solution to account for magneto-resistive effects. Third, the TSA's full capabilities are showcased in non-linear magneto-thermal simulations of several LHC and HL-LHC superconducting magnet models. The full source code, including input files to recreate all simulations, is publicly available.
\end{abstract}

%
\vspace{2pc}

\submitto{\SUST}
%
\maketitle
%
\ioptwocol

\section{Introduction}\label{sec:introduction}

The design and operation of superconducting accelerator magnets necessitate precise thermal management to ensure their stability and performance. During the operation of these magnets, thermal transients, such as those caused by quench events \cite{wilson1983}, can lead to complex thermal behaviours. Their analysis is commonly supported by detailed simulations. The finite element (FE) method can be used for such simulations; see, e.g., \cite{Sirois_2015aa, Bortot_2018aa, Troitino_2019aa, Bortot2020, Stenvall_2023aa}. It offers robust capabilities for modelling the intricate geometries and material properties of superconducting magnets. However, accelerator magnet geometries can consist of superconducting cables with a high ratio of bare conductor thickness to electric insulation thickness \cite{Schnaubelt2023Thermal}, which poses significant challenges for FE discretisation \cite{Driesen2001}.

The thermal gradients across the thin insulation layers must be resolved accurately to ensure the overall validity of the simulation. The accuracy of the FE method depends on the mesh quality \cite[Section 5.3]{Monk_2003aa}. Commonly, a high number of elements is required to create a high-quality mesh in the thin insulation layers \cite{Schnaubelt2023Thermal}. This leads to an increased number of degrees of freedom (DoFs) and, consequently, an increase in computational time. The strong non-linearity of material properties in superconducting accelerator magnets leads to a further rise of required computational resources. Therefore, efficient simulation techniques are desirable.

To address these challenges, this contribution extends the thermal thin-shell approximation (TSA) for the heat diffusion equation proposed in \cite{Schnaubelt2023Thermal} towards real superconducting accelerator magnet geometries. The method is used to compute thermal transients in Large Hadron Collider (LHC) and Hi-Luminosity LHC (HL-LHC) \cite{bruning2024high} superconducting magnets. For two-dimensional (2D) models, the TSA collapses thin thermal insulation surfaces into lines within the FE model, significantly reducing the mesh complexity and the number of required DoFs for high-quality meshes. A similar approach has been proposed in \cite{Bortot_2018aa} using the thermal TSA available in COMSOL Multiphysics\textsuperscript{\textregistered} {\cite{COMSOL}} to replace thin insulation layers between different magnet turns of the winding block. In contrast to \cite{Bortot_2018aa}, instead of replacing only the electric insulation layers between superconducting cables in a block, all insulation domains (e.g., also between different layers) are replaced by the TSA in this work to reduce the number of DoFs even further. Furthermore, while the method of \cite{Bortot_2018aa} enlarges the bare size of the conductor to implement the thin insulation layer, this current work maintains the original bare conductor dimensions.

The presented TSA method is implemented within the open-source Finite Element Quench Simulation tool \texttt{FiQuS} \cite{fiqus}. The \texttt{FiQuS} source code is publicly available at \cite{fiqus} with input files and instructions to recreate all simulations at \cite{multipole-analysis}. The structure of the \texttt{FiQuS} tool and capabilities have already been showcased by the authors in previous works \cite{Vitrano2023, Wozniak2023, Atalay2024, Dular_2024aa}. \texttt{FiQuS} is a free and open-source Python-based tool. It aims to provide the applied superconductivity community with access to quench simulations of superconducting magnets, regardless of their level of expertise in numerical computing. A text-based user interface is provided that separates coil design from numerical aspects, such as suitable FE formulations. It uses \texttt{Gmsh} \cite{gmsh} for computer-aided design, meshing, and post-processing and \texttt{GetDP} \cite{getdp} to compute the FE solution.

This contribution focuses on the \texttt{FiQuS} multipole module dedicated to 2D models of superconducting multipole magnets, which include cos-theta \cite{wilson1983}, block-coil \cite{Sabbi2005}, and common coil magnets \cite{Gupta1997}. The simulations encompass heat generation, thermal diffusion, and quench propagation by solving the heat diffusion equation. To include magneto-resistive effects, a coupling to magnetostatic solutions is considered as well. 

The TSA approach supports multi-layered regions consisting of different temperature-dependent materials with various thicknesses and internal heat sources by considering an internal discretisation \cite{Schnaubelt2023Thermal}. In particular, quench heaters (QHs) \cite{Szeless1998} are modelled with the TSA. Similarly, QHs have been approximated using a TSA in COMSOL Multiphysics\textsuperscript{\textregistered} to analyse the quench propagation in Nb$_3$Sn cables in \cite{Balani_2024aa}. Our TSA implementation supports Dirichlet (imposed temperature), Neumann (imposed heat flux), and Robin (imposed temperature-dependent heat flux) boundary conditions (BCs). This work uses Robin-type BCs are used to approximate the cryogenic cooling of the surrounding cryogenic bath. A coupling of the thermal TSA to a magneto-thermal version has been proposed in \cite{Schnaubelt_2024aa}.

The remainder of this paper is organised as follows: \Sref{sec:tsa} discusses the mathematical model, and briefly addresses the theoretical foundation of the TSA before describing its implementation in \texttt{FiQuS} and the required modifications to the geometry and mesh. \Sref{sec:verification} discusses four numerical case studies to verify and validate the thermal part of the \texttt{FiQuS} multipole module, assess its computational performance, and showcase its features. 

The first two verify the application of the TSA method to superconducting accelerator magnet models by comparison against classical FE models in \texttt{FiQuS} with meshed insulation surfaces. A simple four-cable setup quenched by QHs is considered in \Sref{sec:simple_model} to verify the accuracy and correct implementation of the \texttt{FiQuS} multipole TSA method. A comprehensive simulation of a dipole magnet in \Sref{sec:mbh} shows that the TSA model requires significantly reduced computational effort when compared to a classical FE model for comparable accuracy. \Sref{sec:em_th} considers a magneto-thermal TSA model of the same dipole magnet including QHs and validates the QH delay against measured data. \Sref{sec:four_models} showcases the flexibility of the automated implementation by presenting results for four different LHC and HL-LHC TSA magnet models, including QHs and magneto-resistive effects for magnetostatic fields for time-invariant operating currents. Lastly, \Sref{sec:conclusions} concludes with a discussion on the advantages of the TSA for superconducting magnet quench simulations.

\section{Mathematical Model}\label{sec:tsa}

We are interested in solving the heat equation in the thermal computational domain $\Omega_\text{t}$ that is bounded by $\Gamma_\text{t}$ with outward unit normal vector $\vec{n}_\text{t}$. The boundary is divided into disjunct parts $\Gamma_\text{t} = \Gamma_\text{D} \cup \Gamma_\text{N} \cup \Gamma_\text{R}$ for the Dirichlet, Neumann, and Robin-type BCs. The heat equation reads in strong form: find the temperature $T$ such that
\begin{align}
    - \nabla \cdot \left(\kappa \nabla T \right) + C_\text{V} \partial_t T &= P &&\text{in} \quad \Omega_\text{th}, \\
    T &=T_\text{set} &&\text{on} \quad \Gamma_\text{D},\\
    \vec{n}_\text{t} \cdot ( \kappa \nabla T) &= q && \text{on} \quad \Gamma_\text{N},\\
    \vec{n}_\text{t} \cdot ( \kappa \nabla T) &= h (T_\text{ext} - T)  &&\text{on} \quad \Gamma_\text{R},
 \end{align}
with $\kappa$ the thermal conductivity, $C_\text{V}$ the volumetric heat capacity, $P$ the power density, $T_\text{set}$ the imposed temperature, $q$ the imposed heat flux, $h$ the heat transfer coefficient, and $T_\text{ext}$ the exterior bath temperature. 

\begin{figure}
    \centering
    \begin{tikzpicture}
    \begin{axis}[
        width=0.6\textwidth,
        axis equal image,
        hide axis,
        enlargelimits=false,
        ticks=none,
        legend columns=-1,
        legend style={
            at={(0.5, -0.04)},
            anchor = north, 
            /tikz/every even column/.append style={column sep=0.3cm}
        }
    ] 
            \definecolor{htColor}{RGB}{116,74,36}
            \definecolor{wedgeColor}{RGB}{196,125,59}
            \definecolor{insulationColor}{RGB}{165,165,165}
            
            \addlegendimage{fill=htColor, area legend}
            \addlegendimage{fill=wedgeColor, area legend}
            \addlegendimage{fill=insulationColor, area legend}
            
            \addlegendentry{\footnotesize Half turns $\Omega_\text{ht}$}
            \addlegendentry{\footnotesize Wedges $\Omega_\text{we}$}
            \addlegendentry{\footnotesize Ins. $\Omega_\text{i}$}

            \addplot graphics[xmin=0,xmax=2360,ymin=0,ymax=1307] {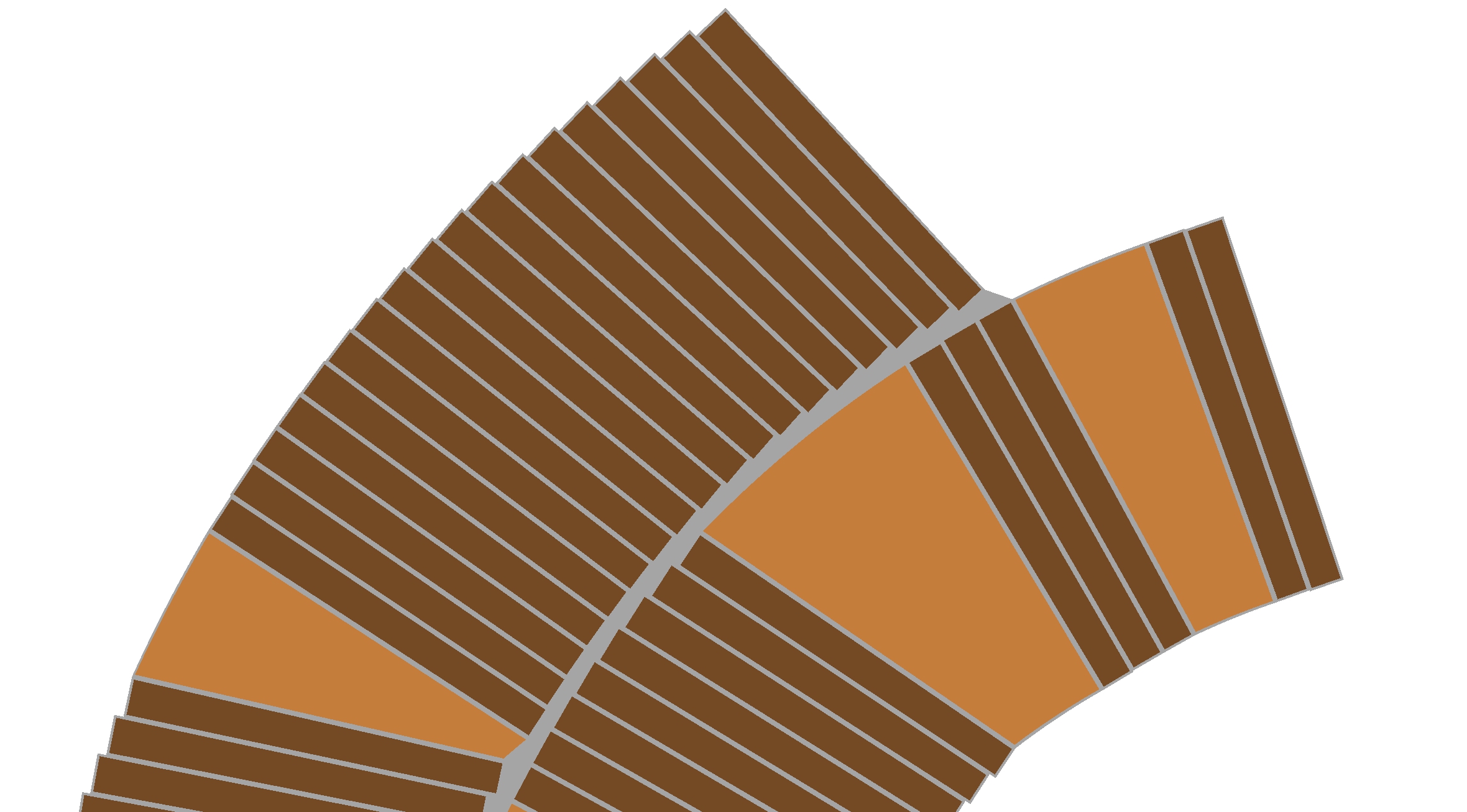};
            
    \end{axis}
\end{tikzpicture}%
    \caption{The thermal computational domain $\Omega_\text{t}$ consists of an electrically insulating domain $\Omega_\text{i}$, metallic wedges $\Omega_\text{we}$ and the half turns $\Omega_\text{ht}$. The insulation domain $\Omega_\text{i}$ is explicitly modelled as a surface. In the TSA model, these surfaces are replaced by lines as shown in \Fref{fig:tsa_description}. The domain is shown for part of the single-aperture dipole MBH \cite{Karppinen_2012aa, Chlachidze_2013aa, Chlachidze_2013ab, Zlobin_2013aa, Zlobin_2015aa, Savary_2015aa, Bordini_2019aa}.}
    \label{fig:computational_domain}
\end{figure}

 As shown in \Fref{fig:computational_domain}, the thermal computational domain $\Omega_\text{t}$ consists of an electrically insulating domain $\Omega_\text{i}$, a region of metallic wedges \cite{Rossi2003} $\Omega_\text{we}$ and a region of magnet bare (half) turns $\Omega_\text{ht}$. Following \cite{Bortot_2018aa}, $\Omega_\text{ht}$ consists of quadrangular half turns $\Omega_{\text{ht}, j}$ such that $\Omega_\text{ht} = \bigcup_j \Omega_{\text{ht}, j}$. Each $\Omega_{\text{ht}, j}$ contains the superconducting filaments, the stabilizer material, and the cable voids, which are potentially filled by an impregnating filling material or liquid helium. Unlike \cite{Bortot_2018aa}, the external insulation is not included in $\Omega_{\text{ht}, j}$ but in  $\Omega_\text{i}$ following \cite{Schnaubelt2023Thermal}. The material composition in $\Omega_{\text{ht},j}$ is represented by a single homogeneous material as discussed in the following.

To address the homogenisation, we introduce the volumetric fractions $f_{i,j}$ for all materials in $\Omega_{\text{ht},j}$ where $i$ denotes the material index. The homogenized heat capacity then reads
\begin{equation}
    C_\text{V}|_{\Omega_{\text{ht}, j}} = \sum_i f_{i,j} C_{\text{V},i, j},
\end{equation}
where $ C_{\text{V},i, j}$ is the volumetric heat capacity of the $i$-th material of the $j$-th half turn $\Omega_{\text{ht}, j}$. In particular, if the voids of the magnet are filled with superfluid helium, the heat capacity of helium is accounted for using the fitted material function from \cite{Verweij_2006aa}. Similarly, the thermal conductivity is given by 
\begin{equation}
    \kappa|_{\Omega_{\text{ht}, j}} = \sum_i f_{i,j} \kappa_{i, j},
\end{equation}
with $\kappa_{i, j}$ the thermal conductivity of the $i$-th material of the $j$-th half turn $\Omega_{\text{ht}, j}$. 

To account for the heat generated by current sharing between the superconducting filaments and the conducting stabilizer, a variant of the Stekly approximation \cite{Stekly_1965aa} is used \cite[Section 18.3]{Russenschuck_2010aa}. In the superconducting state, the operating current $I_{\text{op},j}$ in $\Omega_{\text{ht}, j} $ flows only in the superconducting fraction of the cable, and no Joule heating is considered. If the operating current exceeds the critical current, the excess current flows in the normal conducting matrix. If the critical current is reduced to zero, all current flows in the normal conducting matrix. This simplification avoids solving a non-linear root-finding problem that arises when using the power law for the superconducting fraction (see, e.g., \cite[Section III.D]{Bortot_2020aa}). This approach consequently reduces the computational complexity of the problem. This approximation is appropriate for the simulation of quenching magnets as discussed in \cite[Section 18.3]{Russenschuck_2010aa}, where the exact mathematical form of the transition from fully superconducting to fully quenched cable is of secondary importance. Note that the power law could be used in \texttt{FiQuS} by solving the non-linear root-finding problem with a bisection algorithm, as discussed in \cite{Schnaubelt2023Electromagnetic, Schnaubelt_2024aa}. 

The current sharing approximation is implemented using a polynomial quench state variable $q_\text{s}$ that takes values between 0 and 1. The mathematical expression for $q_\text{s}$ will be given later in this section. The heat source is expressed as
\begin{equation}
    P|_{\Omega_{\text{ht}, j}} = q_\text{s} \, \rho \, \left( \frac{I_{\text{op},j}}{\lvert \Omega_{\text{ht}, j} \rvert} \right)^2,
\end{equation}
with the equivalent homogenized resistivity 
\begin{equation}
    \rho|_{\Omega_{\text{ht}, j}} = \frac{\rho_{\text{stab},j}}{f_{\text{stab},j}},
\end{equation}
with $f_{\text{stab},j}$ the volumetric fraction of the stabilizer in the strands of $\Omega_{\text{ht}, j}$ and $\rho_{\text{stab},j}$ the electrical resistivity of the stabilizer. To define the quench state variable, we introduce the abbreviation
\begin{equation}
    i_{q,j} = 1 - \frac{J_{\text{c},j} \, \lvert \Omega_{\text{ht}, j} \rvert }{\lvert I_{\text{op},j} \rvert} \quad \text{for } I_{\text{op},j} \neq 0.
\end{equation}
Herein, we used the homogenized critical current density
\begin{equation}
   J_{\text{c},j} = f_{\text{SC},j} \, J_\text{c}(\lvert \vec{B} \rvert, T),
\end{equation}
with $J_\text{c}$ the critical current density of the superconducting material, $\vec{B}$ the magnetic flux density and $f_{\text{SC},j}$ the fraction of the superconducting filaments in the $j$\mbox{-}th half turn. Using the previous definitions, we define the quench state variable as a third-order polynomial
\begin{equation}
    q_\text{s}|_{\Omega_{\text{ht}, j}} = \begin{cases}
        0, & \text{if } i_{q,j} \leq 0, \\
        i_{q,j}^2 [-2 i_{q,j} + 3], & \text{if } 0 < i_{q,j} < 1, \\
        1, & \text{if } i_{q,j} = 1.
    \end{cases}
\end{equation}
A third-order polynomial is chosen for $q_\text{s}$ to smoothen the transition between fully superconducting and fully quenched cable which is beneficial for the numerical solver to reach convergence.

The magnetic flux density $\vec{B} = \nabla \times \vec{A}$ is given by the solution to the magnetostatic problem in the domain $\Omega_\text{em}$ with boundary $\Gamma_\text{em}$ and normal vector $\vec{n}_\text{em}$
\begin{align}
    \nabla \times (\nu \nabla \times \vec{A}) &= \vec{J} \quad &&\text{in} \quad \Omega_\text{em},\\
    \vec{n}_\text{em} \times \vec{A} \times \vec{n}_\text{em} &= \vec{0} \quad &&\text{on} \quad \Gamma_\text{em},
\end{align}
with $\vec{A}$ the magnetic vector potential \cite{Lombard_1992aa}, magnetic reluctivity $\nu$ and current density
\begin{equation}
    \vec{J}|_{\Omega_{\text{ht}, j}} = \frac{I_{\text{op},j}}{\lvert \Omega_{\text{ht}, j} \rvert} \vec{e}_z.
\end{equation}
The current density points in the axial direction with unit vector $\vec{e}_z$ since a 2D model in the Cartesian $x$-$y$ plane is considered. The electromagnetic region $\Omega_\text{em}$ contains $\Omega_t$, and additionally the iron yoke with non-linear reluctivity $\nu$, steel collar, and air region, as discussed in \cite{Vitrano2023}. The region $\Omega_\text{em}$ is bounded by $\Gamma_\text{em}$ that is far away from the field source such that the magnetic field is assumed to be zero on $\Gamma_\text{em}$.

In this work, we restrict ourselves to magnetostatic simulations since we only consider the case of time-invariant operating currents in order to focus on the novel thermal transient model. In a future publication, the thermal model discussed in this work will be coupled to an efficient reduced order model that describes the magnetisation and instantaneous power loss in composite superconductors based on \cite{dular2024reducedorderhystereticmagnetization, Dular_2024aa}. 

The heat diffusion and magnetostatic equations are discretised in space using the lowest-order finite element (FE) method. An implicit Euler scheme with an adaptive time step is used for time integration. For the linearisation of the magnetostatic problem, a Newton-Raphson scheme is used while a Picard scheme is used for the heat diffusion equation.

\subsection{Thermal Thin Shell Approximation}
\label{sec:tsa_introduction}

\begin{figure}
    \centering
    \includegraphics[width=\linewidth]{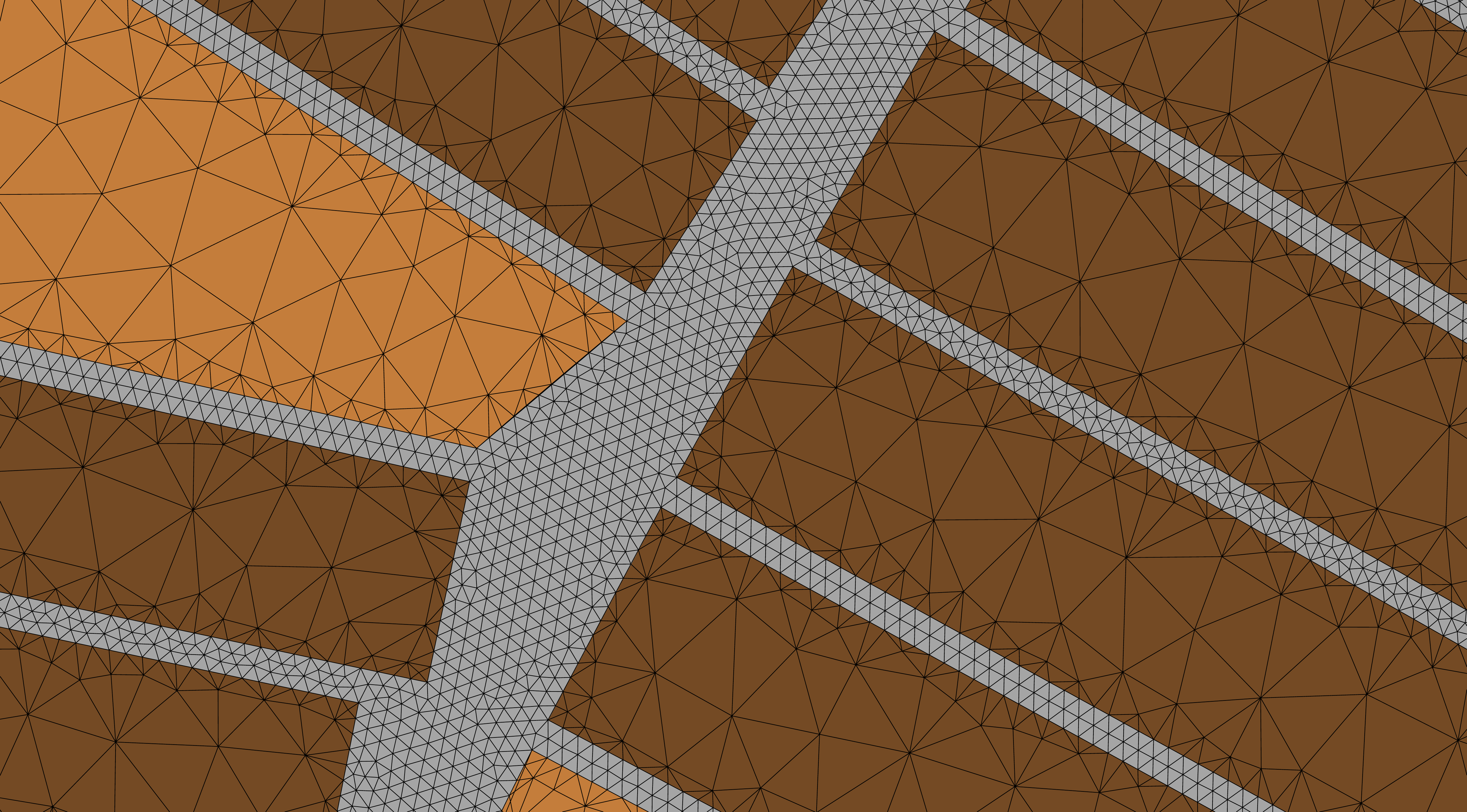}
    \caption{Surface mesh of $\Omega_\text{t}$ of a part of the MBH dipole shown in \Fref{fig:computational_domain}. The fine mesh needed in $\Omega_\text{i}$ forces small elements also in $\Omega_\text{ht}$ and $\Omega_\text{we}$. The mesh for case sur of \Sref{sec:mbh} is shown.}
    \label{fig:sur_mesh_mbh}
\end{figure}

\begin{figure*}[tbh]
    \centering
    \vspace*{-1cm}
    \includegraphics[width=\linewidth]{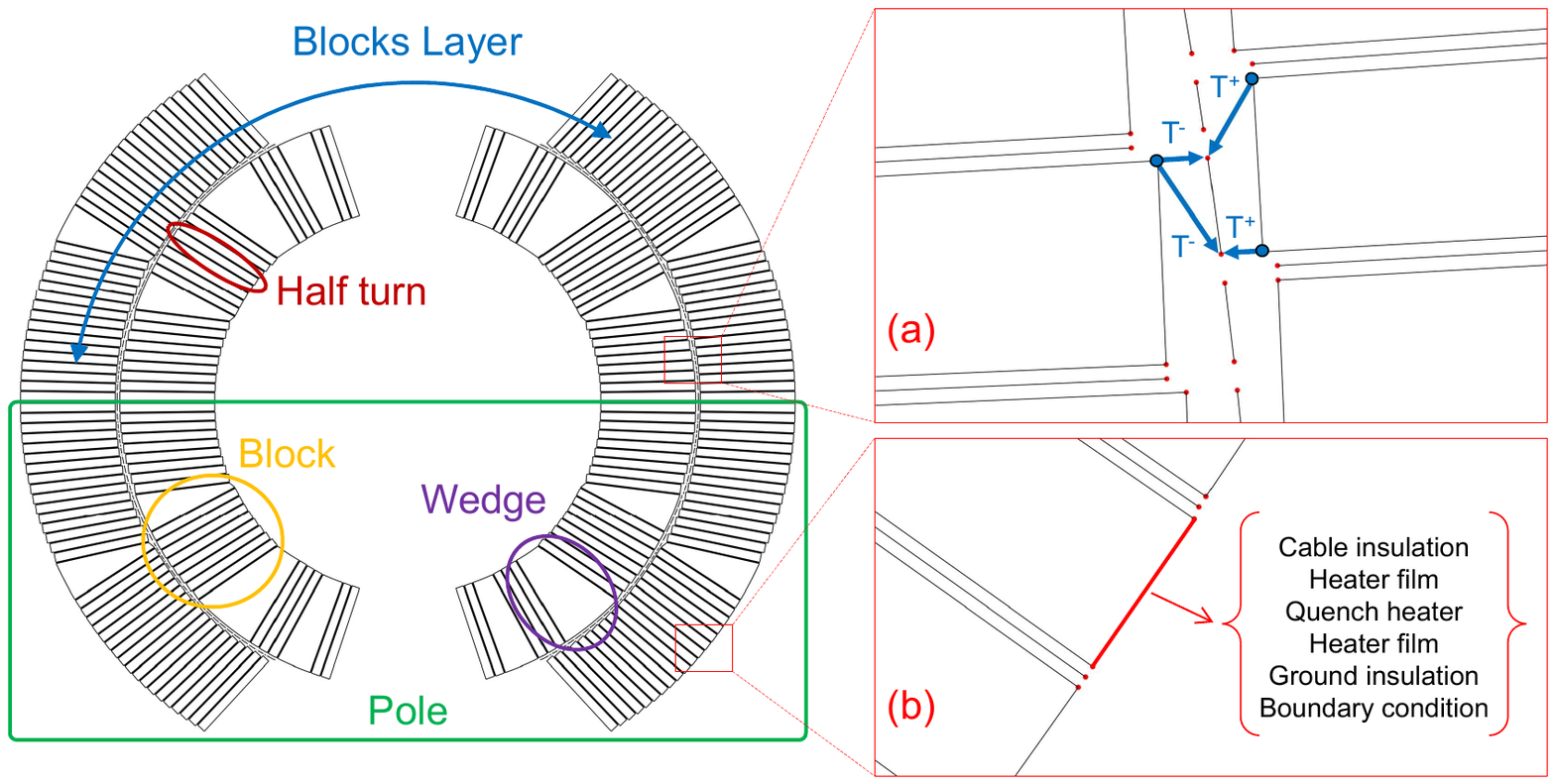}
    \vspace*{-2.8cm}
    \caption{Description of the components of a cos-theta magnet model with close-up views of thin shell lines: a) temperature mapping of the thermally connected regions onto the thin shell line for a coarse mesh with one quadrangular element per conductor; b) breakdown of a realistic combination of layers modelled by an exterior thin shell line marked in red.}
    \label{fig:tsa_description}
\end{figure*}

The previous paragraphs described the classical FE discretisation with surface meshes in the whole thermal domain $\Omega_\text{t}$. In accelerator magnets, the insulation regions $\Omega_\text{i}$ are often thin compared to the half-turn and wedge regions $\Omega_\text{ht}$ and $\Omega_\text{we}$ (see for example \Fref{fig:computational_domain}). Furthermore, the thermal conductivity in $\Omega_\text{ht}$ and $\Omega_\text{we}$ is typically significantly larger than in $\Omega_\text{i}$ leading to larger temperature gradients in $\Omega_\text{i}$ \cite{Schnaubelt2023Thermal}. To compute a globally accurate solution, a high mesh quality is needed in $\Omega_\text{i}$. This is typically achieved by using many small mesh elements; see \Fref{fig:sur_mesh_mbh} for an example. This small mesh also constrains the mesh element size in the neighbouring $\Omega_\text{ht}$ and $\Omega_\text{we}$ regions since conforming meshes are used in the classical FE method. However, since the thermal conductivity of these regions is relatively high, they are approximately isothermal, and few mesh elements are sufficient to capture the small temperature gradient accurately. The aim of the TSA method is to remove the need to mesh $\Omega_\text{i}$ as a surface. This way, coarse meshes of $\Omega_\text{ht}$ and $\Omega_\text{we}$ as shown in \Fref{fig:tsa_description} are straightforward to create. Due to this more flexible mesh, the TSA method can acquire an accurate solution with a significantly reduced number of DoFs.

In this subsection, the application of the thermal TSA for multipole magnet models is discussed. For the TSA, the surface $\Omega_\text{i}$ is replaced by a set of curves $\Gamma_{\text{TSA,i}, j}$ on which the TSA formulation is applied to approximate the behaviour of the solution in $\Omega_\text{i}$. The surface $\Omega_\text{ht}$ is modelled identically for the TSA and the classical FE model, removing the need for scaling factors for the TSA model that were needed in \cite[Section B.2]{Schnaubelt2023Thermal}. All details of the TSA FE formulation, in particular its weak form, can be found in \cite{Schnaubelt2023Thermal}. Furthermore, an extension to magneto-thermal simulations is discussed in \cite{Schnaubelt_2024aa}. 

The TSA uses an internal tensor-product discretisation along the thickness of the thin layer to enable multi-layered regions in a straightforward manner \cite{Schnaubelt2023Thermal}. It also allows for different BCs to model various cooling conditions, heat sources to model QHs and non-linear material properties. Each layer constituting the multi-layer region can be independently discretised across its thickness, allowing for customised resolution of varying temperature gradients. The surface insulation $\Omega_\text{i}$ is approximated by a virtual region $\Omega_{\text{TSA,i}} = \cup_j \Omega_{\text{TSA,i}, j}$ that consists of virtual regions $\Omega_{\text{TSA,i}, j}$ associated with the $j$-th thin shell line (TSL). They are formed by the tensor product of TSLs $\Gamma_{\text{TSA,i}, j}$ and their respective thicknesses $\text{th}_j$, i.e., $\Omega_{\text{TSA,i}, j} = \Gamma_{\text{TSA,i}, j} \times \text{th}_j$. The exterior problem in $\Omega_\text{ht}$ and the interior problem in $\Omega_{\text{TSA,i}}$ need to be connected to ensure the consistency of the overall solution \cite[Section 3.1]{Schnaubelt2023Thermal}.

Each TSL $\Gamma_{\text{TSA,i}, j}$ connects two different surface meshed half turns, referred to as plus and minus sides $\Omega_{\text{ht},\text{TSA},j}^+$ and $\Omega_{\text{ht},\text{TSA},j}^-$, respectively, modelling the heat flux between them. Let us note that the index $j$ refers to the $j$-th TSL, and not the $j$-th half turn. In the model, TSLs are employed to model the multi-layer regions situated between various combinations of half turns, wedges, and domain boundaries (see \Fref{fig:tsa_description}). Specifically, the thin shells replace the insulation regions between half turns (turn-to-turn), block layers (layer-to-layer), coil poles (turn-to-pole), and between conducting elements and domain boundaries (exterior boundaries). \Tref{tab:tsa_types} summarises the layer types present in the different multi-layer regions. Between layers and poles, only surfaces facing each other, i.e., "direct neighbours", are connected thermally as shown in \Fref{fig:tsa_description}(a). The direct heat flux in $\Omega_\text{i}$ across layers and poles to more distant half turns or wedges is not modelled via the TSA since no TSLs are placed between these surface regions. 
This choice is motivated by assuming that the majority of the thermal heat flux is to the closest neighbouring half-turns and wedges, while the 'diagonal' heat flux can be neglected in the model. 

\begin{table}[tbh]
    \begin{indented}
        \centering
        \item[]\begin{tabular}{l | *{4}{c}}
            \br
              & a & b & c & d\\
            \mr
            Short-side cable insulation & & \checkmark & & \checkmark\\
            Long-side cable insulation & \checkmark & & \checkmark & \checkmark\\
            Fish bone \cite{Szeless1998} & & \checkmark & & \\
            Pole separator & & & \checkmark & \\
            Metallic wedge insulation & \checkmark & \checkmark & & \checkmark\\
            QH & & \checkmark & & \checkmark\\
            QH films & & \checkmark & & \checkmark\\
            Ground insulation \cite{Szeless1998} & & & & \checkmark\\
            \br
        \end{tabular}
    \end{indented}
    \caption{Various layer types that can be represented by the TSA: turn-to-turn (a); layer-to-layer (b); turn-to-pole (c); exterior boundaries (d). QHs and the associated insulation films are present only at specific locations.}
    \label{tab:tsa_types}
\end{table}

The internal problem in the TSA domain $\Omega_{\text{TSA,i},j}$ imposes the temperature on the two sides of the TSL, i.e., $T_j^+$ ($T_j^-$) in $\Omega_{\text{ht},\text{TSA},j}^+$ ($\Omega_{\text{ht},\text{TSA},j}^-$), which can be thought of as a BC for the interior problem \cite[Section 3.1]{Schnaubelt2023Thermal}. For the multipole model, the meshes of the plus and minus side and the TSL are not conforming, such that the choice of $T^+$ and $T^-$ is not straightforward. To deal with non-conforming meshes, a mortar TSA approach \cite{Hahn_2025aa} could be used. In this work, to simplify the connection, $T_j^+$ ($T_j^-$) are simply chosen as the temperatures associated to the closest mesh nodes in $\Omega_{\text{ht},j}^+$ ($\Omega_{\text{ht},j}^-$), see \Fref{fig:tsa_description}a. The length of the TSLs is chosen to approximate the contact area between the neighbouring surfaces. The virtual thicknesses of the TSA $\text{th}_j$ is chosen constant per TSL such that the virtual domain $\Omega_{\text{TSA,i}, j}$ is rectangular. The virtual domain $\Omega_{\text{TSA,i}}$ is therefore only an approximation of $\Omega_{\text{i}}$.

In all but one type of multi-layer region, the associated TSL is located between the edges of facing conductors or wedges (see \Fref{fig:tsa_description}a). The exception is the exterior multi-layer regions, where the corresponding TSLs coincide with the exterior edges of the conducting elements (see \Fref{fig:tsa_description}b).
Applying the TSA on the exterior domain boundary allows BCs, such as cooling, to be imposed on the outer boundary of a multi-layer stack adjacent to the meshed region rather than directly on the region itself (see \Fref{fig:tsa_description}b). Furthermore, QHs can be applied by a TSA on the exterior boundary in this way as well.

\subsection{Method Implementation and Usage in \texttt{FiQuS}}\label{sec:implementation}

The necessary steps for the TSA method to work properly are automatically addressed by \texttt{FiQuS}, which takes care of:
\begin{enumerate}[label=(\alph*)]
    \item detecting which conducting elements are thermally directly connected, i.e., find neighbouring elements;
    \item evaluating the extent of contact and the coordinates of the TSL line;
    \item generating TSLs at the geometry level using Gmsh \cite{gmsh};
    \item assembling and assigning the multi-layer regions according to both the magnet insulation information and user-specified parameters; \label{user_inputs}
    \item discretising each layer and meshing the thin shells;
    \item identifying the closest neighbouring nodes for temperature mapping;
    \item generating thin shell physical groups in the mesh via Gmsh;
    \item templating the GetDP-compatible \cite{getdp} input file (i.e., \texttt{pro} file) with the TSA algorithm via Jinja \cite{jinja};
    \item coupling the internal and external problems via GetDP and Jinja for the solution of the thermal problem.
\end{enumerate}

The main user interface of \texttt{FiQuS} is a text-based YAML \cite{Ben2009} input file that simplifies the version control of simulations and, therefore, aids their repeatability. Furthermore, it decouples design parameters regarding geometry and powering schemes from the numerical methods that \texttt{FiQuS} handles in the background. The input files for all simulations in this work and instructions to reproduce them are given in \cite{multipole-analysis}. The results can be reproduced using \texttt{FiQuS} version 2025.1.1 \cite{fiqus}, together with \texttt{CERNGetDP} version 2025.1.2 \cite{cerngetdp}. The temperature- and field-dependent material functions, in particular the $J_\text{c}$ fitting functions, are found in the \texttt{STEAM} material library version 2024.12.2 \cite{steamMaterialLibrary}. The model details for the LHC and HL-LHC models are taken from the \texttt{STEAM} model library version 2025.1.0 \cite{steamModelsLibrary}.

\section{Numerical Case Studies}\label{sec:verification}

In this section, the TSA method implemented in \texttt{FiQuS} is verified against FE models with insulation regions meshed as a surface. We refer to these models as sur models. For the comparison, the average temperature per half-turn
\begin{equation}
    T_{\text{ht}, j} = \int_{\Omega_{\text{ht},j}} \frac{T}{\lvert \Omega_{\text{ht},j} \rvert} \, \text{d}\Omega,
\end{equation}
and the hotspot temperature, i.e., the maximum average half-turn temperature,
\begin{equation}
    T_\text{hotspot} = \text{max}_j \, T_{\text{ht}, j}
\end{equation}
are defined. We have a very fine sur reference model with surface meshed insulation designated as the ref model. Using this ref model, we define the relative error
\begin{equation}
     \varepsilon_{t, \text{ht}, j} = \frac{\lvert T_{\text{ht}, j} - T_{\text{ref}, \text{ht}, j} \rvert}{T_{\text{ref}, \text{ht}, j}},
\end{equation}
the relative hotspot error
\begin{equation}
    \varepsilon_{t, \text{hotspot}} = \frac{\lvert T_\text{hotspot} - T_\text{hotspot, ref} \rvert}{T_\text{hotspot, ref}},
\end{equation}
and the maximum relative error
\begin{equation}
    \varepsilon_{t, \text{max}} = \text{max}_j \, \varepsilon_{t, \text{ht}, j}.
\end{equation}
Last, we define the maximum over time of the latter two errors as $\varepsilon_{\text{hotspot}} = \text{max}_t \, \varepsilon_{t, \text{hotspot}}$ and $\varepsilon_{\text{max}} = \text{max}_t \, \varepsilon_{t, \text{max}}$. 

Four case studies are considered in this section. In the first case study of \Sref{sec:simple_model}, the accuracy and automated implementation of the TSA method is verified by comparison against sur models for a simple arrangement of four superconducting cables heated with QHs. In the second case study of \Sref{sec:mbh}, the computational efficiency of the TSA method for practically relevant problems is highlighted by comparison against sur models for the single-aperture Nb$_{3}$Sn dipole MBH \cite{Karppinen_2012aa, Chlachidze_2013aa, Chlachidze_2013ab, Zlobin_2013aa, Zlobin_2015aa, Savary_2015aa}. For these first two sections, no coupling to a magnetostatic solution is considered. Field-dependent material functions are evaluated for a constant background magnetic flux density. In the third case study of \Sref{sec:em_th}, a magneto-thermal TSA model of MBH with QHs is considered, and the simulated QH delay is compared to measured data. In the fourth case study of \Sref{sec:four_models}, the capabilities of the automated \texttt{FiQuS} TSA implementation are highlighted by showing magneto-thermal simulation results for time-invariant operating currents for four different LHC and HL-LHC magnets.

\subsection{Verification: Four-Cable Model With QHs}\label{sec:simple_model}

\begin{figure*}[tbh]
    \centering    \def\spyx{8.425}
\def\spyy{2.375}

\def\zoomx{8.425}
\def\zoomy{3.75}

\pgfdeclarelayer{background layer}
\pgfdeclarelayer{foreground layer}
\pgfsetlayers{background layer,main,foreground layer}

\begin{tikzpicture}[spy using outlines={width=10cm, height=2cm, every spy on node/.append style={thick}}]

\newlength{\scwidth}
\newlength{\scheight}
\newlength{\insulationthickness}
\newlength{\totalqhlength}
\newlength{\gTenLeft}
\newlength{\kapton}
\newlength{\qh}
\newlength{\gTenRight}
\newlength{\totalqhinsulationthickness}

\setlength{\scwidth}{8.25 cm}
\setlength{\scheight}{1.11525cm}
\setlength{\insulationthickness}{0.075 cm}

\setlength{\gTenLeft}{0.0525 cm}
\setlength{\kapton}{0.03 cm}
\setlength{\qh}{0.01875 cm}
\setlength{\gTenRight}{0.075 cm}

\setlength{\totalqhinsulationthickness}{2\insulationthickness + \gTenLeft + \gTenRight + \qh + 2\kapton}

\fill[TUDa-8b] (0, -\insulationthickness) rectangle (\scwidth, \insulationthickness) ;
\fill[TUDa-1b] (0, 0) rectangle ++(\scwidth, \scheight) node[midway, white, font = \scriptsize] {HT 1} coordinate[midway] (A);
\fill[TUDa-8b] (0, \scheight) rectangle ++(\scwidth, 2\insulationthickness) ;
\fill[TUDa-1b] (0, {\scheight + 2\insulationthickness}) rectangle ++(\scwidth, \scheight) node[midway, white, font = \scriptsize] {HT 2} coordinate[midway] (B);
\fill[TUDa-8b] (0, {2 * \scheight + 2\insulationthickness}) rectangle ++(\scwidth, \insulationthickness) ;

\fill[TUDa-8b] ({\scwidth + \totalqhinsulationthickness}, -\insulationthickness) rectangle ++(\scwidth, \insulationthickness) ;
\fill[TUDa-1b] ({\scwidth + \totalqhinsulationthickness}, 0) rectangle ++(\scwidth, \scheight) node[midway, white, font = \scriptsize] {HT 3} coordinate[midway] (C);
\fill[TUDa-8b] ({\scwidth + \totalqhinsulationthickness}, \scheight) rectangle ++(\scwidth, 2\insulationthickness) ;
\fill[TUDa-1b] ({\scwidth + \totalqhinsulationthickness}, {\scheight + 2\insulationthickness}) rectangle ++(\scwidth, \scheight) node[midway, white, font = \scriptsize] {HT 4} coordinate[midway] (D);
\fill[TUDa-8b] ({\scwidth + \totalqhinsulationthickness}, {2 * \scheight + 2\insulationthickness}) rectangle ++(\scwidth, \insulationthickness) ;

\fill[TUDa-8b] (-\insulationthickness, 0) rectangle ++(\insulationthickness, \scheight) ;
\fill[TUDa-8b] (-\insulationthickness, {\scheight + 2\insulationthickness}) rectangle ++(\insulationthickness, \scheight) ;

\fill[TUDa-8b] ({2\scwidth + \totalqhinsulationthickness}, 0) rectangle ++(\insulationthickness, \scheight) ;
\fill[TUDa-8b] ({2\scwidth + \totalqhinsulationthickness}, {\scheight + 2\insulationthickness}) rectangle ++(\insulationthickness, \scheight) ;

\fill[TUDa-8b] (\scwidth, 0) rectangle ++(\insulationthickness, \scheight) ;
\fill[TUDa-8b] (\scwidth, {\scheight + 2\insulationthickness}) rectangle ++(\insulationthickness, \scheight) ;
\fill[TUDa-8b] ({\scwidth + \totalqhinsulationthickness - \insulationthickness}, 0) rectangle ++(\insulationthickness, \scheight) ;
\fill[TUDa-8b] ({\scwidth + \totalqhinsulationthickness - \insulationthickness}, {\scheight + 2\insulationthickness}) rectangle ++(\insulationthickness, \scheight) ;

\fill[TUDa-11c] (0, -\insulationthickness - \gTenRight) rectangle ++(\scwidth, \gTenRight) ;
\fill[TUDa-3d] (0, {-\insulationthickness - \gTenRight - 2\kapton}) rectangle ++(\scwidth, 2\kapton) ;

\fill[TUDa-11c] (\scwidth + \totalqhinsulationthickness, {2\scheight + 3\insulationthickness}) rectangle ++(\scwidth, \gTenRight) ;
\fill[TUDa-3d] (\scwidth + \totalqhinsulationthickness, {2\scheight + 3\insulationthickness + \gTenRight}) rectangle ++(\scwidth, 2\kapton) ;

\fill[TUDa-11c] (\scwidth + \insulationthickness, 0) rectangle ++(\gTenLeft, \scheight) ;
\fill[TUDa-3d] (\scwidth + \insulationthickness + \gTenLeft, 0) rectangle ++(\kapton, \scheight) ;
\fill[TUDa-9b] (\scwidth + \insulationthickness + \gTenLeft + \kapton, 0) rectangle ++(\qh, \scheight) ;
\fill[TUDa-3d] (\scwidth + \insulationthickness + \gTenLeft + \kapton + \qh, 0) rectangle ++(\kapton, \scheight) ;
\fill[TUDa-11c] (\scwidth + \insulationthickness + \gTenLeft + 2\kapton + \qh, 0) rectangle ++(\gTenRight, \scheight) ;

\fill[TUDa-11c] (\scwidth + \insulationthickness, \scheight + 2\insulationthickness) rectangle ++(\gTenLeft, \scheight) ;
\fill[TUDa-3d] (\scwidth + \insulationthickness + \gTenLeft, \scheight + 2\insulationthickness) rectangle ++(\kapton, \scheight) ;
\fill[TUDa-9b] (\scwidth + \insulationthickness + \gTenLeft + \kapton, \scheight + 2\insulationthickness) rectangle ++(\qh, \scheight) ;
\fill[TUDa-3d] (\scwidth + \insulationthickness + \gTenLeft + \kapton + \qh, \scheight + 2\insulationthickness) rectangle ++(\kapton, \scheight) ;
\fill[TUDa-11c] (\scwidth + \insulationthickness + \gTenLeft + 2\kapton + \qh, \scheight + 2\insulationthickness) rectangle ++(\gTenRight, \scheight) ;

\fill[TUDa-11c] (2\scwidth + \insulationthickness + \totalqhinsulationthickness, 0) rectangle ++(\gTenLeft, \scheight) ;
\fill[TUDa-3d] (2\scwidth + \insulationthickness + \totalqhinsulationthickness + \gTenLeft, 0) rectangle ++(\kapton, \scheight) ;
\fill[TUDa-9b] (2\scwidth + \insulationthickness + \totalqhinsulationthickness + \gTenLeft + \kapton, 0) rectangle ++(\qh, \scheight) ;
\fill[TUDa-3d] (2\scwidth + \insulationthickness + \totalqhinsulationthickness + \gTenLeft + \kapton + \qh, 0) rectangle ++(\kapton, \scheight) ;
\fill[TUDa-11c] (2\scwidth + \insulationthickness + \totalqhinsulationthickness + \gTenLeft + 2\kapton + \qh, 0) rectangle ++(\gTenRight, \scheight) ;

\fill[TUDa-11c] (2\scwidth + \insulationthickness + \totalqhinsulationthickness, \scheight + 2\insulationthickness) rectangle ++(\gTenLeft, \scheight) ;
\fill[TUDa-3d] (2\scwidth + \insulationthickness + \totalqhinsulationthickness + \gTenLeft, \scheight + 2\insulationthickness) rectangle ++(\kapton, \scheight) ;
\fill[TUDa-9b] (2\scwidth + \insulationthickness + \totalqhinsulationthickness + \gTenLeft + \kapton, \scheight + 2\insulationthickness) rectangle ++(\qh, \scheight) ;
\fill[TUDa-3d] (2\scwidth + \insulationthickness + \totalqhinsulationthickness + \gTenLeft + \kapton + \qh, \scheight + 2\insulationthickness) rectangle ++(\kapton, \scheight) ;
\fill[TUDa-11c] (2\scwidth + \insulationthickness + \totalqhinsulationthickness + \gTenLeft + 2\kapton + \qh, \scheight + 2\insulationthickness) rectangle ++(\gTenRight, \scheight) ;

\spy [black,magnification=20,spy connection path={\draw[thick] (tikzspyonnode) -- (tikzspyinnode);}] on (\spyx, \spyy) in node at (\zoomx, \zoomy);

 \begin{pgfonlayer}{foreground layer}    
    \draw[] (5.7, 4.3125) -- (4.3125, 4.3125);
    \draw[] (5.7, 4.3125) -- (5.7, 3.5625);
    \node[fill = white, font = \scriptsize, align=center] at (5.7, 4.3125) {Cable\\ ins.};

    \draw[] (6.95, 4.3125) -- (6.95, 3.5625);
    \node[fill = white, font = \scriptsize, align=center] at (6.95, 4.3125) {G10\\ layer};

    \draw[] (8.26, 4.3125) -- (8.26, 3.5625);
    \node[fill = white, font = \scriptsize, align=center] at (8.26, 4.3125) {QH};
    
    \draw[] (9.1, 4.3125) -- (8.75, 3.5625);
    \node[fill = white, font = \scriptsize, align=center] at (9.1, 4.3125) {Kapton\\ film};

    \node[font = \scriptsize, white] at (4.2, 3.3) {HT 2};
    \node[font = \scriptsize, white] at (12.7, 3.3) {HT 4};
\end{pgfonlayer}

\path let \p1 = (A) in coordinate (a) at (7.87,\y1);
\draw[white] (a) node[white, font = \scriptsize, align=center, left]
{QH 1} -- ++(0.55, 0);

\path let \p1 = (B) in coordinate (b) at (7.87,\y1);
\draw[white] (b) node[white, font = \scriptsize, align=center, left]
{QH 2} -- ++(0.55, 0);

\path let \p1 = (C) in coordinate (c) at (16.475,\y1);
\draw[white] (c) node[white, font = \scriptsize, align=center, left]
{QH 3} -- ++(0.55, 0);

\path let \p1 = (D) in coordinate (d) at (16.475,\y1);
\draw[white] (d) node[white, font = \scriptsize, align=center, left]
{QH 4} -- ++(0.55, 0);

\end{tikzpicture}
    \caption{Schematic illustration of the four-conductor model with a close-up view of a layer-to-layer multi-layer region. The proportions between layer thicknesses are in accordance with the actual model.}
    \label{fig:4_conductor}
\end{figure*}
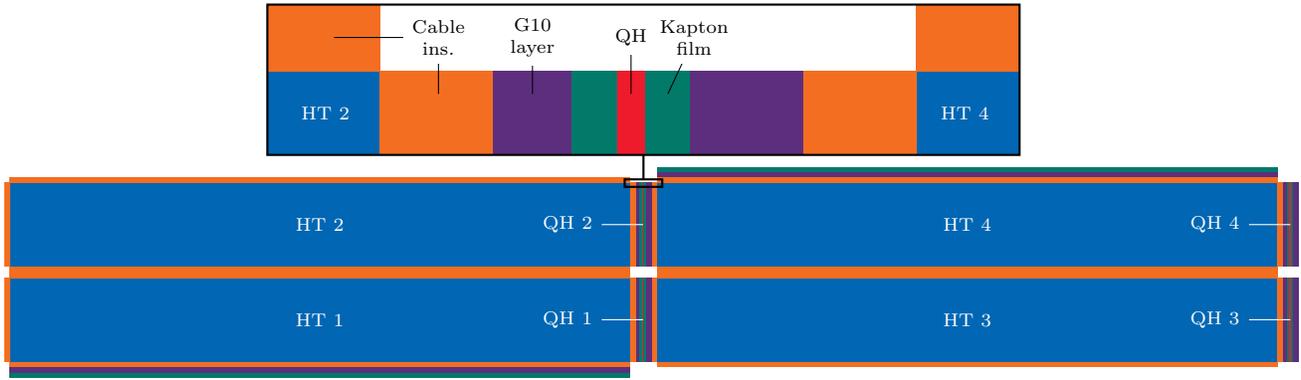

The first case study considers a simple arrangement of four superconducting cables as shown in \Fref{fig:4_conductor}. It consists of a two-layer configuration of superconductor blocks, with one block per layer and two cables per block. The left side of the first layer (left)  and the right side of the second layer (right) represent the inner and outer surfaces of the coil, respectively. Cables with {Nb-Ti/Cu} strands with characteristic geometrical and material composition representative of the LHC conductors are chosen. The four bare conductors are surrounded by an arbitrarily chosen combination of insulation layers. This is done to demonstrate that any multi-layer region can be accurately modelled using TSA, regardless of its position and composition. Specifically:
\begin{itemize}
    \item each bare Nb-Ti/Cu half turn (HT) is surrounded by a cable insulation layer;
    \item the layers are separated by a region consisting of several insulation layers and two quench heaters (QH 1 and QH 2, see the close-up view in \Fref{fig:4_conductor});
    \item the outer side of the coil is covered by a similar sequence of layers, which includes two additional quench heaters (QH 3 and QH 4);
    \item two multi-layer insulation groups are placed on the lower and upper sides of the first and last half turn (HT 1 and HT 4), respectively.
\end{itemize}

The insulation layers are positioned such that heat flux is only considered between half turns that face each other directly. In particular, the insulation layers do not cover the edges of the half turns such that there is no diagonal heat flux, e.g., from SC 1 to SC 4. For this setup, the TSA assumptions of negligible diagonal heat flux discussed in \Sref{sec:tsa_introduction} and rectangular domains $\Omega_{\text{TSA,i}, j}$ are fulfilled exactly. Therefore, the TSA and models with surface meshed insulation are equivalent. This choice was taken to verify the correctness of the automated multi-layer TSA implementation against an equivalent model. The TSA models are built fully automated from a YAML file, while the sur models are built manually. This is because the algorithm creating the sur model geometry cannot create insulation layers made from multiple materials since their creation as a surface would require complicated geometry creation and meshing steps. In particular, QHs cannot be automatically created for the sur models. To this end, the four-cable model is quite simple to make a manually built sur model feasible.

\begin{figure*}[tb]
    \centering
    \begin{subfigure}{.49\linewidth}
        \centering
        \begin{tikzpicture}
    \begin{axis}[
        width=\textwidth,
        axis equal image,
        hide axis,
        enlargelimits=false,
        point meta min = 11.7,
        point meta max = 304,
        colormap name=viridis,
        colorbar horizontal,
        colorbar style={
            title=Temperature $\left(\si{\kelvin}\right)$ at $t \equal \SI{360}{\milli \second}$,
            title style={
                at={(0.5,-2)},
                anchor=north,
                font=\footnotesize
            },
            scaled x ticks=false,
            xtick style={draw=none},
            xticklabel style = {font=\footnotesize},
            xtick={11.6, 304},
            at={(0.5,-0.05)},
            anchor=north,
            width = .5\textwidth,
        },
        colorbar/width=0.25cm,
    ] 
        \addplot graphics[xmin=0,xmax=32.32,ymin=0,ymax=8.56] {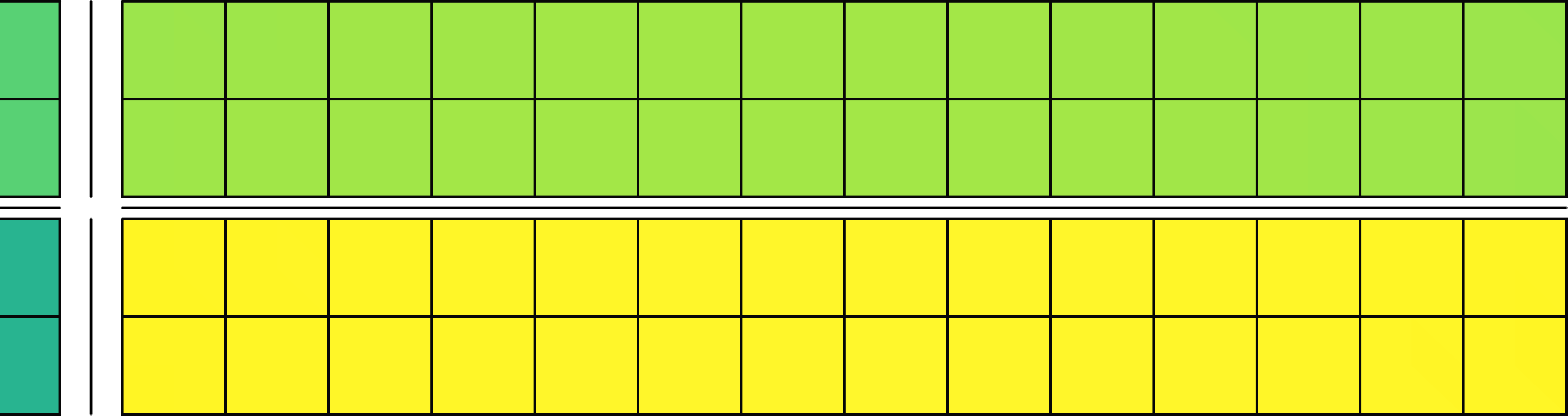};
    \end{axis}
\end{tikzpicture}%
    \caption{TSA model.}%
    \end{subfigure}
    \hfill
    \begin{subfigure}{.49\linewidth}
            \centering
            \begin{tikzpicture}
    \begin{axis}[
        width=\textwidth,
        axis equal image,
        hide axis,
        enlargelimits=false,
        point meta min = 11.7,
        point meta max = 304,
        colormap name=viridis,
        colorbar horizontal,
        colorbar style={
            title=Temperature $\left(\si{\kelvin}\right)$ at $t \equal \SI{360}{\milli \second}$,
            title style={
                at={(0.5,-2)},
                anchor=north,
                font=\footnotesize
            },
            scaled x ticks=false,
            xtick style={draw=none},
            xticklabel style = {font=\footnotesize},
            xtick={11.6, 304},
            at={(0.5,-0.075)},
            anchor=north,
            width = .5\textwidth,
        },
        colorbar/width=0.25cm,
    ] 
        \addplot graphics[xmin=0,xmax=32.32,ymin=0,ymax=8.56] {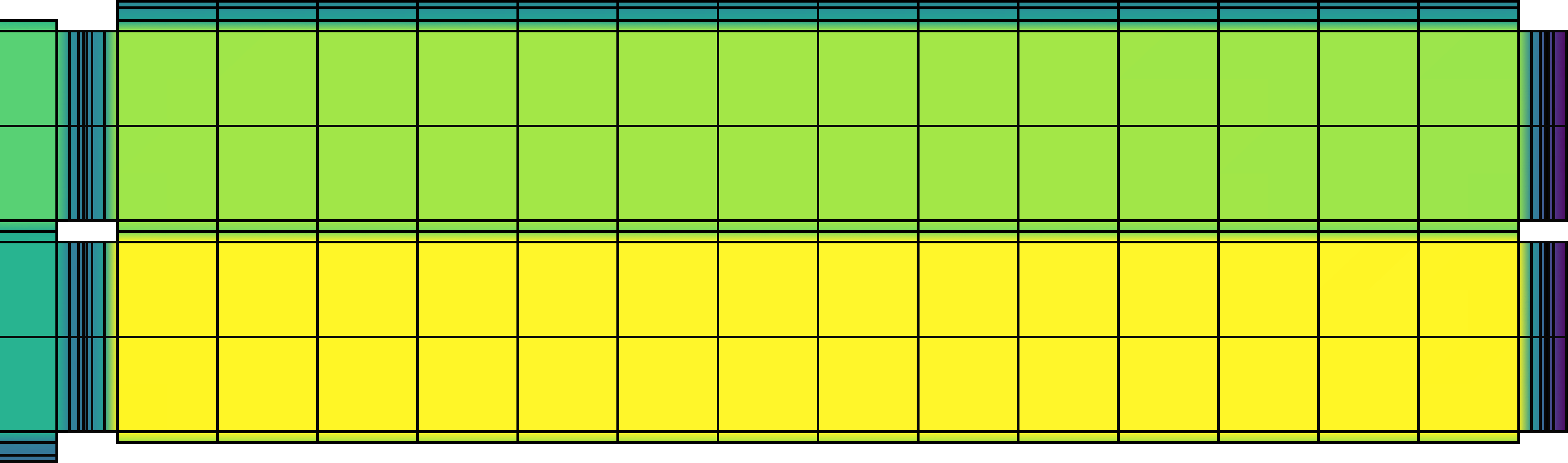};
    \end{axis}
\end{tikzpicture}%
            \caption{Sur model.}%
    \end{subfigure}
    \caption{The final temperature map and FE mesh of two models for the four-cable case are shown; the same colour scale is used for both models. The coarsest sur model mesh of \Fref{fig:mesh_sensitivity_simple_model} is shown for clear visualization. The superconducting half turns are always meshed with exactly this quadrangular pattern while the number of mesh elements across the insulation thickness is increased. The refinement level of the TSA is not visible since an internal discretization is used \cite{Schnaubelt2023Thermal}. Due to the high thermal conductivity of the cable compared to the insulation, the half turns are approximately isothermal. For example, the maximum temperature of HT 3 at $t = \SI{360}{\milli \second}$ is approximately \SI{304}{\kelvin} while the minimum temperature is around \SI{299}{\kelvin}.}
    \label{fig:simple_meshes}
\end{figure*}

\Fref{fig:simple_meshes} shows the structured quadrangular mesh. Inside the half turn, an identical mesh is used for the TSA and sur model. For the insulation layer, the number of TSA discretization layers is chosen to be identical to the number of surface mesh elements of the sur model. Furthermore, the Gaussian points of the quadrangular surface of the sur model and the TSA line elements \cite{Schnaubelt2023Thermal} are chosen at the same location to enable an equivalent situation for both models leading to excellent agreement between them.

\begin{figure}[tbh]
    \centering
    \pgfplotstableread[col sep=comma,]{data_4cond/time_steps.csv}\timeSteps
\pgfplotstableread[col sep=comma,]{data_4cond/T_time_ref.csv}\Tref
\pgfplotstableread[col sep=comma,]{data_4cond/T_time_tsa.csv}\Ttsa

\pgfplotstablecreatecol[copy column from table={\timeSteps}{0}]{timeSteps}{\Tref}
\pgfplotstablecreatecol[copy column from table={\timeSteps}{0}]{timeSteps}{\Ttsa}

\begin{tikzpicture}

    \begin{axis}[cycle list={[
        colormap/viridis, 
        colors of colormap={0, 300, 600, 900, 0, 300, 600, 900}]},
        width=.47\textwidth,
        height=6cm,
        xlabel={Time (\si{\milli \second})},
        ylabel={HT temperature $T_{\text{ht},j}$ (\si{\kelvin})}, 
        ylabel near ticks, 
        scaled x ticks=false, 
        legend columns=2, 
        legend style={font=\footnotesize, at={(0.01, 0.99)}, anchor=north west}, 
        ticklabel style = {font=\scriptsize},
        legend style={font=\scriptsize},
        label style={font=\small},
        legend image post style={mark indices={}}, 
        scaled ticks=false, 
        ytick scale label code/.code={}, 
        legend cell align={left},
        legend image post style={mark indices={}}, 
        x filter/.code={\pgfmathparse{#1*1000}\pgfmathresult},
        ymin = -15,
        ymax = 325,
        xmin = -25, 
        xmax = 380
         ]

      \addplot+[
                mark=none,
                thick
            ] 
         table[
            x = timeSteps,
            y index = 0
        ] 
        {\Tref};%

    \addplot+[
                mark=none,
                thick
            ] 
         table[
            x = timeSteps,
            y index = 1
        ] 
        {\Tref};%

    \addplot+[
                mark=none,
                thick
            ] 
         table[
            x = timeSteps,
            y index = 2
        ] 
        {\Tref};%

    \addplot+[
                mark=none,
                thick
            ] 
         table[
            x = timeSteps,
            y index = 3
        ] 
        {\Tref};%

    \addplot+[
    mark=none,
    mymark={o}{solid, opacity=1}, 
    opacity=0,
    thick
        ] 
     table[
        x = timeSteps,
        y index = 0
    ] 
    {\Ttsa};%

    \addplot+[
    mark=none,
    mymark={o}{solid, opacity=1}, 
    opacity=0,
    thick
    ] 
     table[
        x = timeSteps,
        y index = 1
    ] 
    {\Ttsa};%

    \addplot+[
    mark=none,
    mymark={o}{solid, opacity=1}, 
    opacity=0,
    thick
    ] 
     table[
        x = timeSteps,
        y index = 2
    ] 
    {\Ttsa};%

    \addplot+[
    mark=none,
    mymark={o}{solid, opacity=1}, 
    opacity=0,
    thick
    ] 
     table[
        x = timeSteps,
        y index = 3
    ] 
    {\Ttsa};%

    \addplot[black, dashed, mark=none] coordinates {(0.01, -30) (0.01, 400)};

    \addplot[black, dotted, mark=none] coordinates {(0.05, -30) (0.05, 400)};
         
    \legend{sur HT 1, sur HT 2, sur HT 3, sur HT 4, TSA HT 1, TSA HT 2, TSA HT 3, TSA HT 4, QH 2 on, QH 3 on}
    
    \end{axis}%
\end{tikzpicture}
    \caption{The temperature evolution of the four half turns is compared between the sur and TSA models. The legend reports the activation time of the QHs and the average temperature of the bare superconducting half turns. The solution is shown for the finest mesh element thickness considered in \Fref{fig:mesh_sensitivity_simple_model}.}
    \label{fig:comparison_simple_model}
\end{figure}
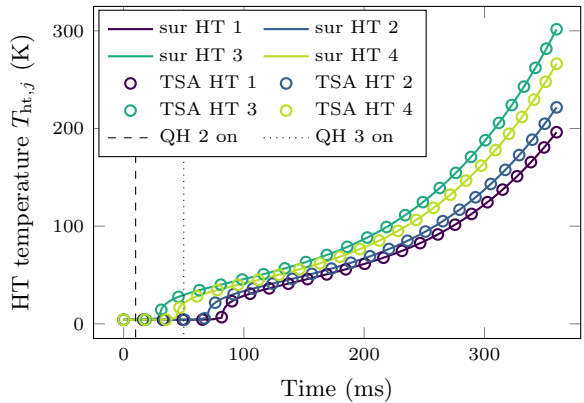

An exponentially decaying current is considered in the stainless steel strips of QH 2 and QH 3 starting from $t = \SI{10}{\milli \second}$ and $t = \SI{50}{\milli \second}$, respectively. The generated Joule heating raises the temperature of the conductors. Subsequently, the conductors quench and begin generating volumetric heat, propagating the quench to the neighbouring half-turns. Cryogenic cooling with a constant heat transfer coefficient is applied on the exterior side of QH 3 and QH 4; all other boundaries are adiabatic, to check the implementation of this BC.  The simulation is run until $t = \SI{360}{\milli \s}$ where $T_\text{hotspot} > \SI{300}{\kelvin}$ as shown in \Fref{fig:simple_meshes}. \Fref{fig:comparison_simple_model} shows the average temperature of the half turns $T_{\text{ht}, j}$ at each time step throughout the transient for both the sur and TSA models with both models in excellent agreement as expected.

\begin{figure}[tbh]
    \centering\pgfplotstableread[col sep=comma,]{data_4cond/rel_error_mesh_sizes.csv}\meshSizes
\pgfplotstableread[col sep=comma,]{data_4cond/rel_error_mesh_tsa.csv}\meshErrorTSA
\pgfplotstableread[col sep=comma,]{data_4cond/rel_error_mesh_ref.csv}\meshErrorRef

\pgfplotstablecreatecol[copy column from table={\meshSizes}{0}]{meshSize}{\meshErrorTSA}
\pgfplotstablecreatecol[copy column from table={\meshErrorTSA}{0}]{TSAError}{\meshErrorTSA}

\pgfplotstablecreatecol[copy column from table={\meshSizes}{0}]{meshSize}{\meshErrorRef}
\pgfplotstablecreatecol[copy column from table={\meshErrorRef}{0}]{TSAError}{\meshErrorRef}

\begin{tikzpicture}

    \begin{axis}[cycle list={[
        colormap/viridis, 
        colors of colormap={0, 600}]},
        width=.47\textwidth,
        height=5cm,
        xlabel={Insulation mesh element thickness (\si{\meter})},
        ylabel={Max. rel. err. $\varepsilon_{\text{max}}$ (\%)}, 
        ylabel near ticks, 
        scaled x ticks=false, 
        legend columns=1, 
        legend style={font=\footnotesize, at={(0.01, 0.99)}, anchor=north west}, 
        ticklabel style = {font=\scriptsize},
        legend style={font=\scriptsize},
        label style={font=\small},
        legend image post style={mark indices={}}, 
        scaled ticks=false, 
        ytick scale label code/.code={}, 
        legend cell align={left},  
        ymode=log,
        xmode=log
         ]

      \addplot+[
            mark=o,
            thick
            ] 
         table[
            x = meshSize,
            y = TSAError
        ] 
        {\meshErrorRef};%
         
     \addplot+[
            mark=x,
            thick, 
            dashed,
            mark options=solid
            ] 
         table[
            x = meshSize,
            y = TSAError
        ] 
        {\meshErrorTSA};%
    \legend{sur, TSA}
    \end{axis}%
\end{tikzpicture}%
    \caption{A mesh sensitivity study of the sur and TSA models of the four-cable model is shown. Errors are computed with respect to a ref model with insulation mesh element thickness of \SI{1}{\micro \meter}. The mesh inside the half turns is kept constant, while the number of mesh elements across the insulation thickness is increased.}
    \label{fig:mesh_sensitivity_simple_model}
\end{figure}
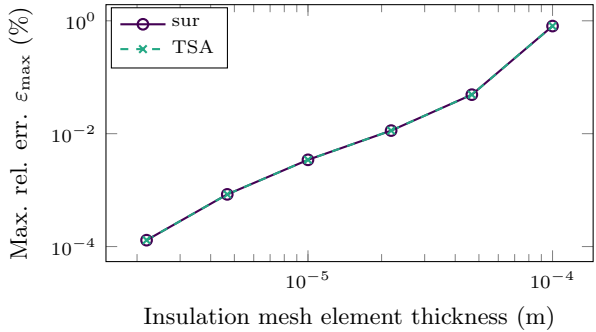

A mesh refinement study is conducted for a more thorough and systematic comparison. As shown in \Fref{fig:simple_meshes}, the solution is quasi-isothermal inside the superconducting half turns with large temperature gradients across the insulation layer. Hence, only a few elements are needed inside the half turns for an accurate solution. The mesh inside the half turn is kept identical to the one shown in \Fref{fig:simple_meshes} for the mesh sensitivity study. Across the thickness of the insulation, more elements are needed to calculate the temperature gradient accurately. \Fref{fig:mesh_sensitivity_simple_model} shows the maximum relative error $\epsilon_\text{max}$ as a function of the insulation mesh element thickness. The relative error decreases monotonically with increased number of insulation mesh elements. The TSA and sur models achieve the same accuracy for the same discretization level as expected. This verifies the correct implementation of the TSA multi-layer algorithm.

\subsection{Verification: Cos-Theta Multipole Magnet}\label{sec:mbh} 

The second case study in this section considers the single-aperture Nb$_{3}$Sn dipole MBH \cite{Karppinen_2012aa, Chlachidze_2013aa, Chlachidze_2013ab, Zlobin_2013aa, Zlobin_2015aa, Savary_2015aa}, whose geometry was previously shown in \Fref{fig:tsa_description}. As shown in \Fref{fig:computational_domain}, the insulation layer is a single Kapton surface that surrounds each conducting element with varying thicknesses. Multi-layer regions are omitted here since their discretization would lead to prohibitive computational cost in the sur model given the large number of half turns. This limitation highlights another disadvantage of the sur models using the classical FE method without TSA: complex multi-layer configurations are computationally unfeasible for fully meshed models, except in very simple cases such as the one shown in the previous section.

A thermal transient is simulated in which certain conductors are forced to the normal conducting state by setting their critical current density to zero. The conductors selected for quenching are those adjacent to the QH strips installed in MBH. The simulations are run until $T_\text{hotspot} > \SI{300}{\kelvin}$. As one of the main quantities of interest, the comparison will focus on the hotspot temperature $T_\text{hotspot}$ and its relative error $\epsilon_\text{hotspot}$. 

Again, a mesh sensitivity study was performed by successively refining the mesh inside the insulation layer. Then, the coarsest TSA and the coarsest sur model are identified that have a hotspot error $\epsilon_\text{hotspot}$ below \SI{0.5}{\percent}. This procedure allows a fair comparison of the computational effort needed to achieve comparable accuracy in an important quantity of interest for the sur and TSA model. The mesh of the sur model is shown in \Fref{fig:sur_mesh_mbh}, and the half turn and wedge mesh of the TSA model is shown in \Fref{fig:tsa_description}. In particular, a coarse mesh of one single quadrangular mesh element is used per half turn and wedge in the TSA model.

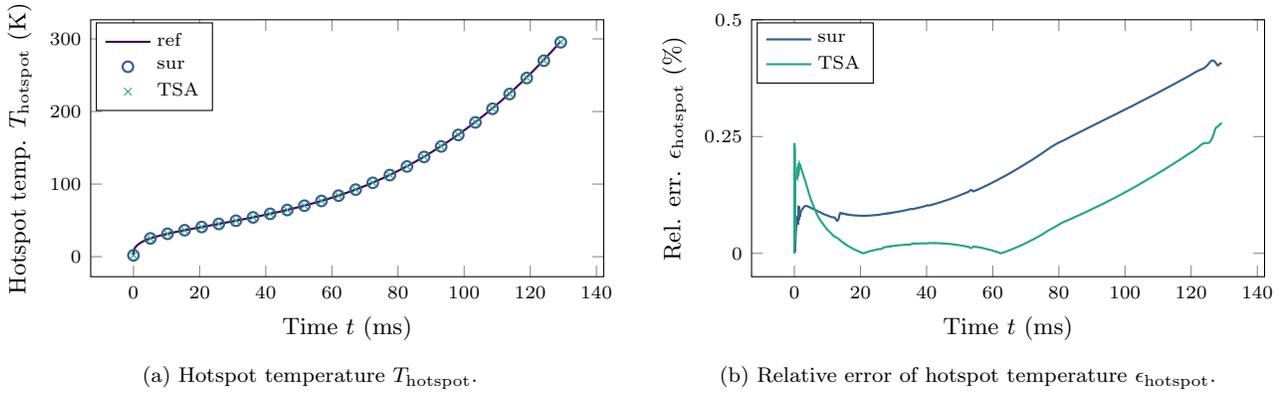
\begin{figure*}[tb]
    \centering
    \begin{subfigure}[b]{.49\linewidth}
        \centering
        \begin{tikzpicture}

    \begin{axis}[cycle list={[
        colormap/viridis, 
        colors of colormap={0, 300, 600}]},
        width=\textwidth,
        height=5cm,
        xlabel={Time $t$ (\si{\milli \second})},
        ylabel={Hotspot temp. $T_\text{hotspot}$ (\si{\kelvin})}, 
        ylabel near ticks, 
        scaled x ticks=false, 
        legend columns=1, 
        legend style={font=\footnotesize, at={(0.01, 0.99)}, anchor=north west}, 
        ticklabel style = {font=\scriptsize},
        legend style={font=\scriptsize},
        label style={font=\small},
        legend image post style={mark indices={}}, 
        scaled ticks=false, 
        ytick scale label code/.code={}, 
        legend cell align={left},
        x filter/.code={\pgfmathparse{#1*1000}\pgfmathresult}
         ]

      \addplot+[
            thick
        ] 
         table[
            col sep=comma,
            x expr = \thisrowno{0},
            y expr = \thisrowno{1}
        ] 
        {data_mbh/T_hotspot_ref.csv};%

        \addplot+[
            thick,
            mark=o,
            only marks,
            mark indices = {1, 100, 200, 300, 400, 500, 600, 700, 800, 900, 1000, 1100, 1200, 1300, 1400, 1500, 1600, 1700, 1800, 1900, 2000, 2100, 2200, 2300, 2400, 2499}
            ]
         table[
            col sep=comma,
            x expr = \thisrowno{0},
            y expr = \thisrowno{1}
        ] 
        {data_mbh/T_hotspot_REF_1.00E-04.csv};%

    \addplot+[
            mark=x,
            only marks,
            mark indices = {1, 100, 200, 300, 400, 500, 600, 700, 800, 900, 1000, 1100, 1200, 1300, 1400, 1500, 1600, 1700, 1800, 1900, 2000, 2100, 2200, 2300, 2400, 2499}
            ] 
         table[
            col sep=comma,
            x expr = \thisrowno{0},
            y expr = \thisrowno{1}
        ] 
        {data_mbh/T_hotspot_TSA_0.0001.csv};%
    \legend{ref, sur, TSA}
    \end{axis}%
\end{tikzpicture}%
        \caption{Hotspot temperature $T_\text{hotspot}$.}
    \end{subfigure}
    \hfill
    \begin{subfigure}[b]{.49\linewidth}
        \centering
        \begin{tikzpicture}

    \begin{axis}[cycle list={[
        colormap/viridis, 
        colors of colormap={300, 600}]},
        width=\textwidth,
        height=5cm,
        xlabel={Time $t$ (\si{\milli \second})},
        ylabel={Rel. err. $\epsilon_\text{hotspot}$ (\%)}, 
        ylabel near ticks, 
        scaled x ticks=false, 
        legend columns=1, 
        legend style={font=\footnotesize, at={(0.01, 0.99)}, anchor=north west}, 
        ticklabel style = {font=\scriptsize},
        legend style={font=\scriptsize},
        label style={font=\small},
        legend image post style={mark indices={}}, 
        scaled ticks=false, 
        ytick scale label code/.code={}, 
        legend cell align={left},  
        x filter/.code={\pgfmathparse{#1*1000}\pgfmathresult},
        y filter/.code={\pgfmathparse{#1*100}\pgfmathresult},
        ytick={0, 0.25, 0.5},
        ymax = 0.5,
         ]

        \addplot+[
            thick,
            ] 
         table[
            col sep=comma,
            x expr = \thisrowno{0},
            y expr = \thisrowno{1}
        ] 
        {data_mbh/err_hotspot_REF_1.00E-04.csv};%

    \addplot+[
        thick,
            ] 
         table[
            col sep=comma,
            x expr = \thisrowno{0},
            y expr = \thisrowno{1}
        ] 
        {data_mbh/err_hotspot_TSA_0.0001.csv};%

    \legend{sur, TSA}%
    \end{axis}%
\end{tikzpicture}%
        \caption{Relative error of hotspot temperature $\epsilon_\text{hotspot}$.}
    \end{subfigure}
    \caption{Hotspot temperature for the MBH model quenched by locally setting $J_\text{c}$ to zero for all three models. The relative error in the hotspot temperature of the sur and TSA models compared to the fine ref model is shown as well.}
    \label{fig:hotspot_mbh}
\end{figure*}

The hotspot temperature of the three models and the relative error of the {sur} and {TSA} models with respect to the {ref} model are shown in \Fref{fig:hotspot_mbh}. The three models are in good agreement with errors in the hotpot temperature below \SI{0.5}{\percent} as required. Furthermore, the number of DoFs and computational time for the three models are shown in \Tref{tab:verification_results}. To reach the same $\epsilon_\text{hotspot}$, the {TSA} is more than 38 times faster than the {sur} model, highlighting its computational efficiency due to significantly fewer DoFs (4608 vs 206037). Last, it can be noticed that the time for the assembly of the linear systems $t_\text{a}$ is significantly higher than for the solution of the linear systems $t_\text{s}$. While this is not surprising for 2D simulations, the difference is significant. It is mostly due to the material fitting functions that contain lengthy mathematical expressions and need to be called for each Gaussian point in the assembly stage. For a further reduction of computational time, the evaluation of the material functions should thus be made more computationally efficient, potentially by a parallelisation approach or by pre-computing a lookup table.

\begin{table}[tbh]
    \begin{indented}
        \centering
        \item[]\begin{tabular}{l | *{3}{c}}
            \br
             & ref & sur & TSA\\
            \mr
            $N_\text{dof}$  & 2611556 & 206037 & 4608\\
            $t_\text{a}$ in \si{\second} & 121518 & 42064 & 1268\\
            $t_\text{s}$ in \si{\second} & 63135 & 7830 & 27\\
            $t_\text{a} + t_\text{s}$ in \si{\second} & 184653 & 49894 & 1295 \\
            \br
        \end{tabular}
    \end{indented}
    \caption{The number of DoFs $N_\text{dof}$, total time spent on system assembly $t_\text{a}$, total time spent on system solution $t_\text{s}$, and the sum of the latter two are shown for the three different MBH models.}
    \label{tab:verification_results}
\end{table}

\subsection{Validation: QH Delay for Magneto-Thermal Cos-Theta Multipole Magnet Model}\label{sec:em_th}

Having verified the accuracy and efficiency of the TSA approach, the third case study will validate the \texttt{FiQuS} QH TSA implementation by comparing the QH delay to measured data for the MBH dipole. The QH delay is the time between activation of the QH and the point in time where the transport current density exceeds the homogenized critical current density in one point of a superconducting half turn. This validation includes coupling to magnetostatic solutions described in detail in \cite{Vitrano2023}. The magnetostatic simulations include the iron yoke, incorporating non-linear $B$-$H$ curves. Since the mesh requirements for electromagnetic simulations differ significantly from those of thermal simulations, different meshes are used. The magnetostatic problem is solved first, yielding the magnetic flux density $\vec{B}$. The latter is then evaluated once at the Gaussian points of the thermal mesh, stored, and reused for all subsequent assembly steps. This approach allows the use of non-linear material properties that depend on both temperature and magnetic flux density.

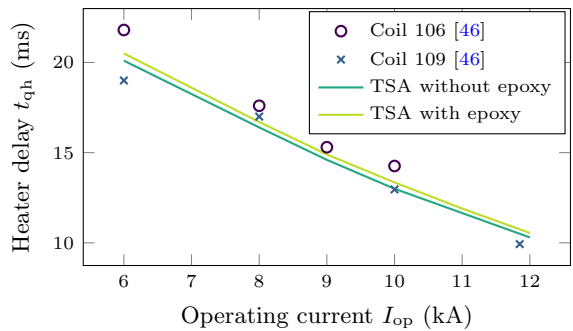
\begin{figure}[t]
    \centering
    \begin{tikzpicture}

    \begin{axis}[cycle list={[
        colormap/viridis, 
        colors of colormap={0, 300, 600, 900}]},
        width=.47\textwidth,
        height=5cm,
        xlabel={Operating current $I_\text{op}$ (\si{\kilo \ampere})},
        ylabel={Heater delay $t_\text{qh}$ (\si{\milli \second})}, 
        ylabel near ticks, 
        scaled x ticks=false, 
        legend columns=1, 
        legend style={font=\footnotesize, at={(0.99, 0.99)}, anchor=north east}, 
        ticklabel style = {font=\scriptsize},
        legend style={font=\scriptsize},
        label style={font=\small},
        legend image post style={mark indices={}}, 
        scaled ticks=false, 
        legend cell align={left},  
         ]

       \addplot+[
            only marks,
            mark=o,
            thick,
            x filter/.code={\pgfmathparse{\pgfmathresult*0.001}},
            y filter/.code={\pgfmathparse{\pgfmathresult*1000}}
        ] 
         coordinates {
            (6000, 21.8e-3)
            (8000, 17.6e-3)
            (9000, 15.3e-3)
            (10000, 14.26e-3)
    };

   \addplot+[
        only marks,
        mark=x,
        thick,
        x filter/.code={\pgfmathparse{\pgfmathresult*0.001}},
        y filter/.code={\pgfmathparse{\pgfmathresult*1000}}
        ] 
     coordinates {
        (6000, 19e-3)
        (8000, 17e-3)
        (10000, 12.96e-3)
        (11850, 9.93e-3)
    };

      \addplot+[
            thick,
            x filter/.code={\pgfmathparse{\pgfmathresult*0.001}},
            y filter/.code={\pgfmathparse{\pgfmathresult*1000}}
        ] 
         coordinates {
            (6000, 0.0201)
            (7000, 0.01825)
            (8000, 0.0164)
            (9000, 0.0146)
            (10000, 0.013)
            (11000, 0.01165)
            (12000, 0.0103)
    };

      \addplot+[
            thick,
            x filter/.code={\pgfmathparse{\pgfmathresult*0.001}},
            y filter/.code={\pgfmathparse{\pgfmathresult*1000}}
    ] 
         coordinates {
            (6000, 0.0205)
            (7000, 0.0186)
            (8000, 0.0167)
            (9000, 0.0149)
            (10000, 0.01335)
            (11000, 0.0119)
            (12000, 0.01055)
    };
    
    \legend{Coil 106 \cite{Izquierdo-Bermudez_2016aa}, Coil 109 \cite{Izquierdo-Bermudez_2016aa}, {TSA without epoxy}, {TSA with epoxy}}
    \end{axis}%
\end{tikzpicture}
    \caption{QH delay as a function of the operating current computed with the \texttt{FiQuS} TSA model and measured as reported in \cite[Figure 4]{Izquierdo-Bermudez_2016aa}. \texttt{FiQuS} TSA results are shown with and without considering the \SI{15}{\micro \meter} epoxy adhesive layer between the polyimide film and the stainless steel heater strips as reported in \cite[Section 9.3.1.4]{Bordini_2019aa}.}
    \label{fig:qh_delay}
\end{figure}

For all simulations of this section, cryogenic cooling BCs using a temperature-dependent heat transfer coefficient following \cite{Verweij_2006aa} are applied on the outer boundary of the coil in radial direction. Adiabatic BCs are applied on all other boundaries. The QH design parameters are taken from \cite{Izquierdo-Bermudez_2016aa} and lead to a peak initial QH current of \SI{150}{\ampere}. FiQuS solves for the current and the resulting power in the QH strips. \Fref{fig:qh_delay} shows the QH delay as a function of the operating current compared to measurements extracted from \cite[Figure 4]{Izquierdo-Bermudez_2016aa}. \texttt{FiQuS} TSA results are shown with and without considering the \SI{15}{\micro \meter} epoxy adhesive layer between the polyimide film and the stainless steel heater strips as reported in \cite[Section 9.3.1.4]{Bordini_2019aa}. The TSA results agree with the measurements for the operating currents considered and the epoxy adhesive shows a negligible influence on the QH heater delay.

\subsection{Magneto-Thermal Simulations of Four LHC and HL-LHC Magnets}\label{sec:four_models}

\begin{figure*}[t]
    \centering
    \begin{subfigure}{.24\textwidth}
        \centering
        \begin{tikzpicture}
    \begin{axis}[
        width=1.45\linewidth,
        axis equal image,
        hide axis,
        enlargelimits=false,
        point meta min = 0,
        point meta max = 8.65,
        colormap name=viridis,
        colorbar horizontal,
        colorbar style={
            title=Magn. Flux Density $\left(\si{\tesla}\right)$,
            title style={
                at={(0.5,-2)},
                anchor=north,
                font=\footnotesize
            },
            scaled x ticks=false,
            xtick style={draw=none},
            xticklabel style = {font=\footnotesize},
            xtick={0, 8.65},
            at={(0.5,-0.025)},
            anchor=north,
            width = .5\textwidth,
        },
        colorbar/width=0.25cm,
    ] 
    
        \addplot graphics[xmin=0,xmax=1300,ymin=0,ymax=1298] {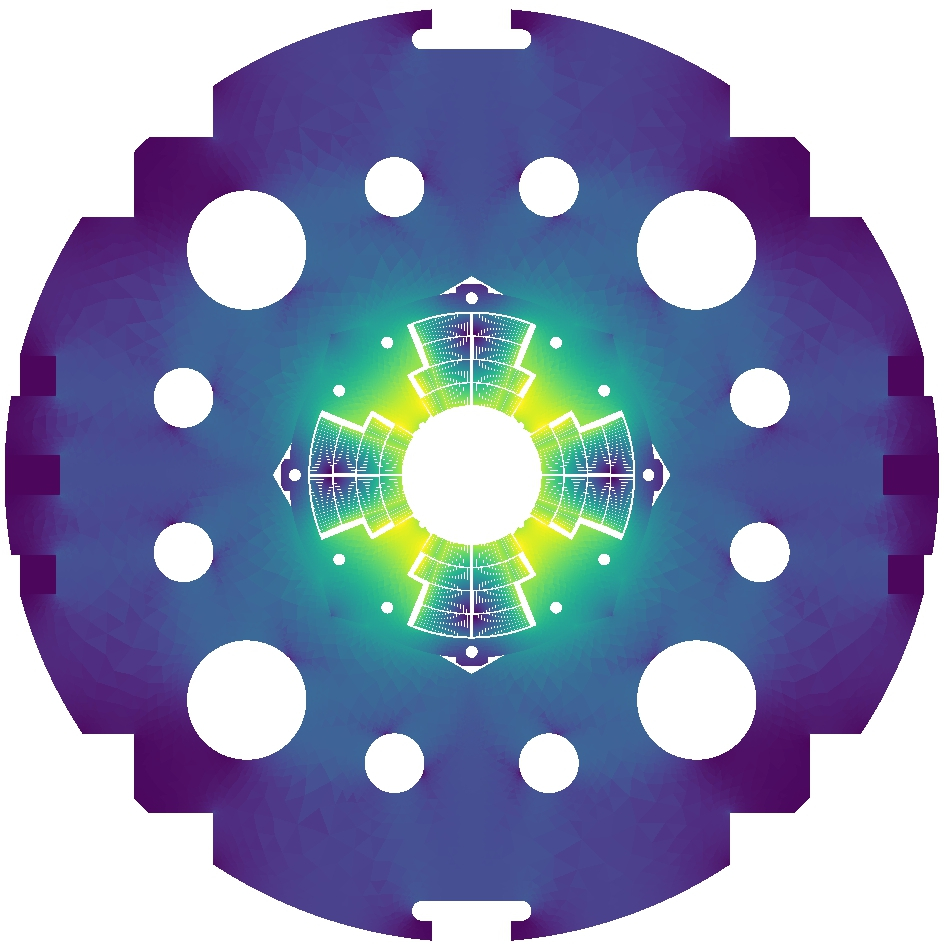};
    \end{axis}
\end{tikzpicture}%
        \caption{MQXA flux density.}
        \label{fig:mqxa_em}
    \end{subfigure}
    \hspace*{-0.5cm}
    \begin{subfigure}{.24\textwidth}
        \centering  
        \begin{tikzpicture}
    \begin{axis}[
        width=1.45\linewidth,
        hide axis,
        axis equal image,
        enlargelimits=false,
        point meta min = 1.9,
        point meta max = 302,
        colormap/plasma,
        colorbar horizontal,
        colorbar style={
            title=Temperature $\left(\si{\kelvin}\right)$,
            title style={
                at={(0.5,-2)},
                anchor=north,
                font=\footnotesize
            },
            scaled x ticks=false,
            xtick style={draw=none},
            xticklabel style = {font=\footnotesize},
            xtick={1.9, 302},
            at={(0.5,-0.025)},
            anchor=north,
            width = .5\textwidth,
        },
        colorbar/width=0.25cm,
    ] 
        \addplot graphics[xmin=0,xmax=1291,ymin=0,ymax=1291] {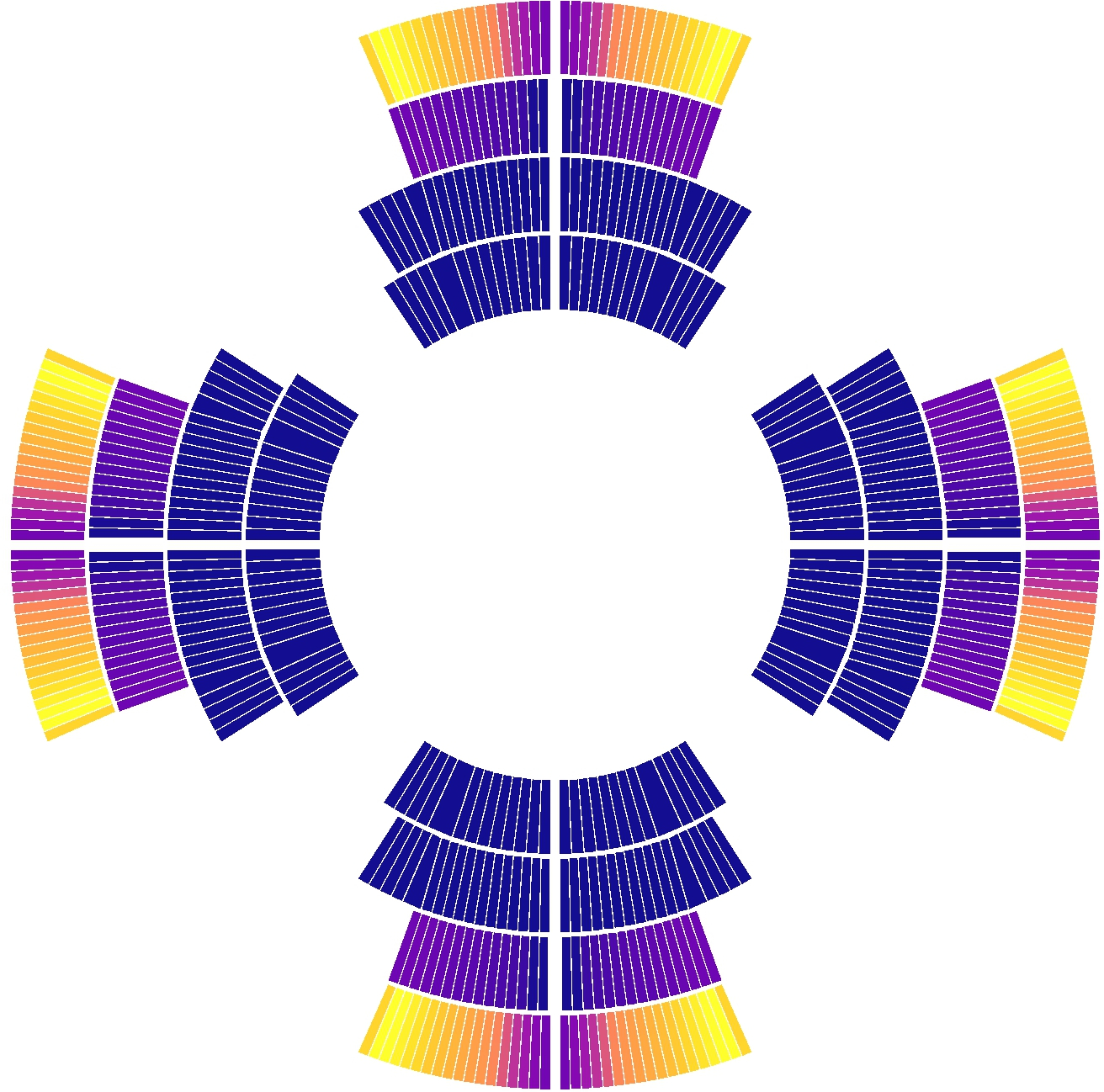};
    \end{axis}
\end{tikzpicture}%
        \caption{MQXA temperature.}%
        \label{fig:mqxa_th}%
    \end{subfigure}
    \hspace*{-0.2cm}
    \begin{subfigure}{.24\textwidth}
        \centering
        \begin{tikzpicture}
    \begin{axis}[
        width=1.45\textwidth,
        axis equal image,
        hide axis,
        enlargelimits=false,
        point meta min = 0,
        point meta max = 8.62,
        colormap name=viridis,
        colorbar horizontal,
        colorbar style={
            title=Magn. Flux Density $\left(\si{\tesla}\right)$,
            title style={
                at={(0.5,-2)},
                anchor=north,
                font=\footnotesize
            },
            scaled x ticks=false,
            xtick style={draw=none},
            xticklabel style = {font=\footnotesize},
            xtick={0, 8.62},
            at={(0.5,-0.025)},
            anchor=north,
            width = .5\textwidth,
        },
        colorbar/width=0.25cm,
    ] 
    
        \addplot graphics[xmin=0,xmax=1314,ymin=0,ymax=1311] {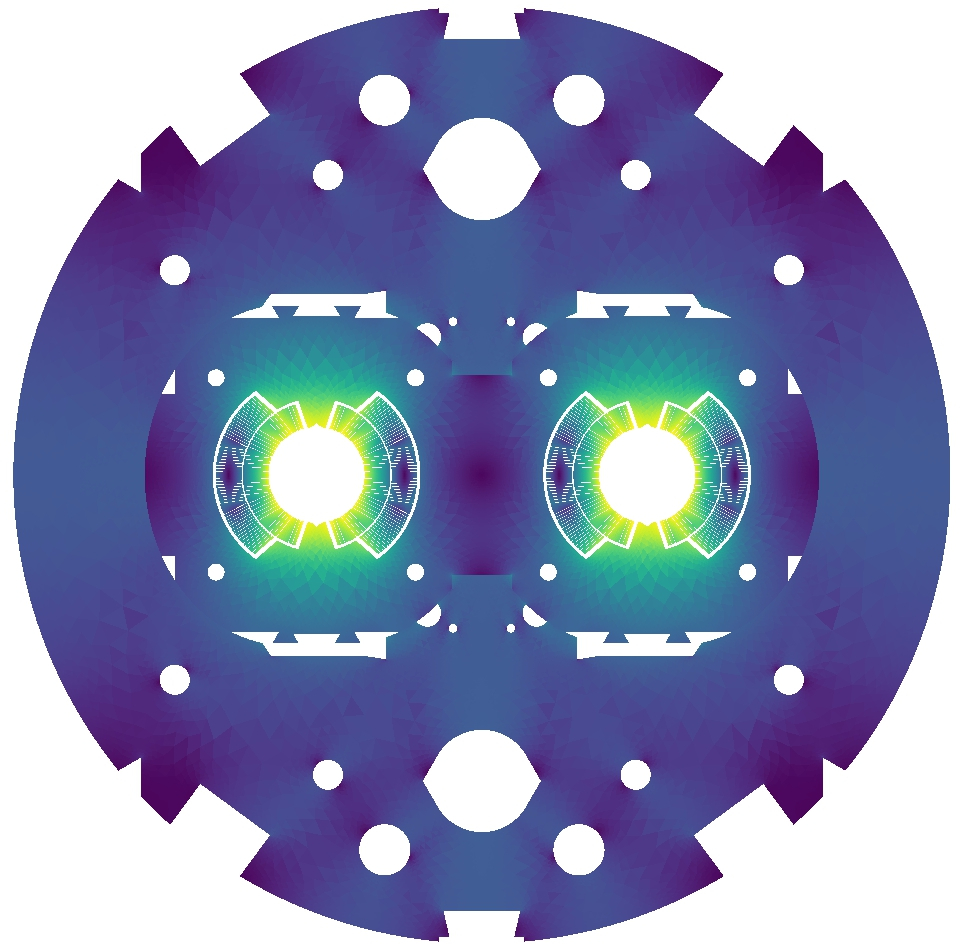};
    \end{axis}
\end{tikzpicture}%
        \caption{MB flux density.}
        \label{fig:mb_em}
    \end{subfigure}
    \hspace*{0.1cm}
    \begin{subfigure}{.24\textwidth}
        \centering
        \begin{tikzpicture}
    \begin{axis}[
        width=1.45\textwidth,
        axis equal image,
        hide axis,
        enlargelimits=false,
        point meta min = 1.9,
        point meta max = 304,
        colormap/plasma,
        colorbar horizontal,
        colorbar style={
            title=Temperature $\left(\si{\kelvin}\right)$,
            title style={
                at={(0.5,-2)},
                anchor=north,
                font=\footnotesize
            },
            scaled x ticks=false,
            xtick style={draw=none},
            xticklabel style = {font=\footnotesize},
            xtick={1.9, 304},
            at={(0.5,-0.025)},
            anchor=north,
            width = .5\textwidth,
        },
        colorbar/width=0.25cm,
    ] 
        \addplot graphics[xmin=0,xmax=1617,ymin=0,ymax=1277] {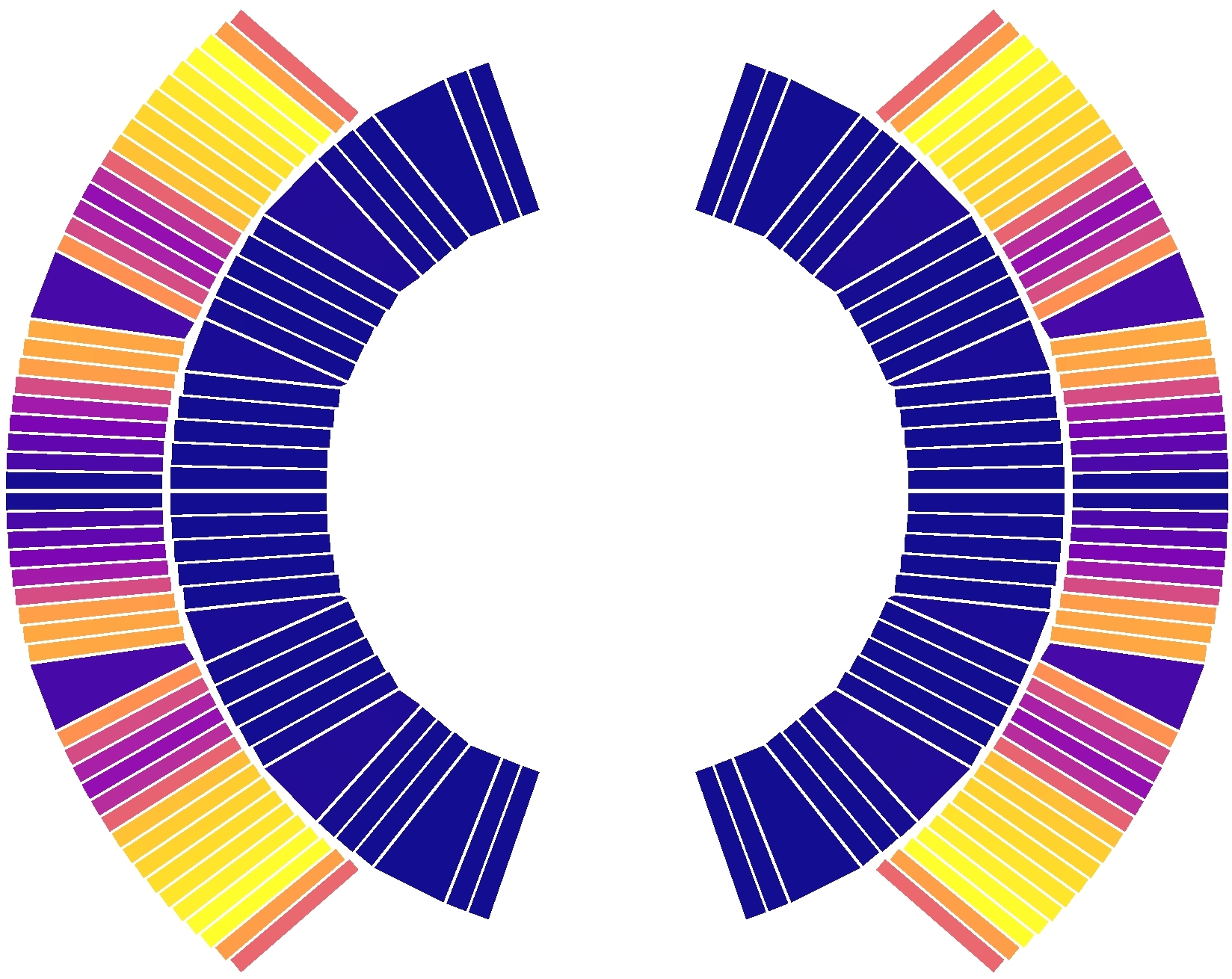};
    \end{axis}
\end{tikzpicture}%
        \caption{MB temperature.}
        \label{fig:mb_th}
    \end{subfigure}
    \\[.5em]
    \begin{subfigure}{.24\linewidth}
        \centering
        \begin{tikzpicture}
    \begin{axis}[
        width=1.45\textwidth,
        axis equal image,
        hide axis,
        enlargelimits=false,
        point meta min = 0,
        point meta max = 11.4,
        colormap name=viridis,
        colorbar horizontal,
        colorbar style={
            title=Magn. Flux Density $\left(\si{\tesla}\right)$,
            title style={
                at={(0.5,-2)},
                anchor=north,
                font=\footnotesize
            },
            scaled x ticks=false,
            xtick style={draw=none},
            xticklabel style = {font=\footnotesize},
            xtick={0, 11.4},
            at={(0.5,-0.025)},
            anchor=north,
            width = .5\textwidth,
        },
        colorbar/width=0.25cm,
    ] 
        \addplot graphics[xmin=0,xmax=1311,ymin=0,ymax=1315] {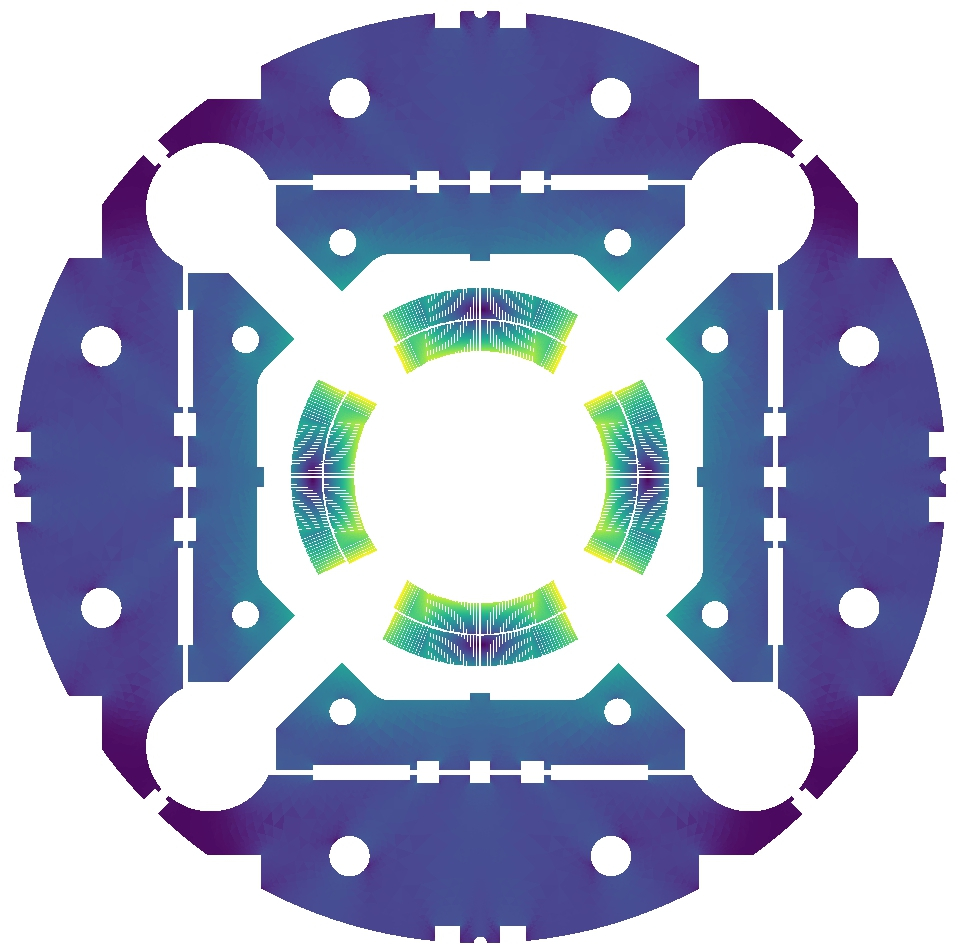};
    \end{axis}
\end{tikzpicture}%
        \caption{MQXF flux density.}
        \label{fig:mqxf_em}
    \end{subfigure}
    \hspace*{-0.5cm}
    \begin{subfigure}{.24\linewidth}
        \centering
        \begin{tikzpicture}
    \begin{axis}[
        width=1.45\textwidth,
        axis equal image,
        hide axis,
        enlargelimits=false,
        point meta min = 2.7,
        point meta max = 302,
        colormap/plasma,
        colorbar horizontal,
        colorbar style={
            title=Temperature $\left(\si{\kelvin}\right)$,
            title style={
                at={(0.5,-2)},
                anchor=north,
                font=\footnotesize
            },
            scaled x ticks=false,
            xtick style={draw=none},
            xticklabel style = {font=\footnotesize},
            xtick={2.7, 302},
            at={(0.5,-0.025)},
            anchor=north,
            width = .5\textwidth,
        },
        colorbar/width=0.25cm,
    ] 
        \addplot graphics[xmin=0,xmax=1297,ymin=0,ymax=1301] {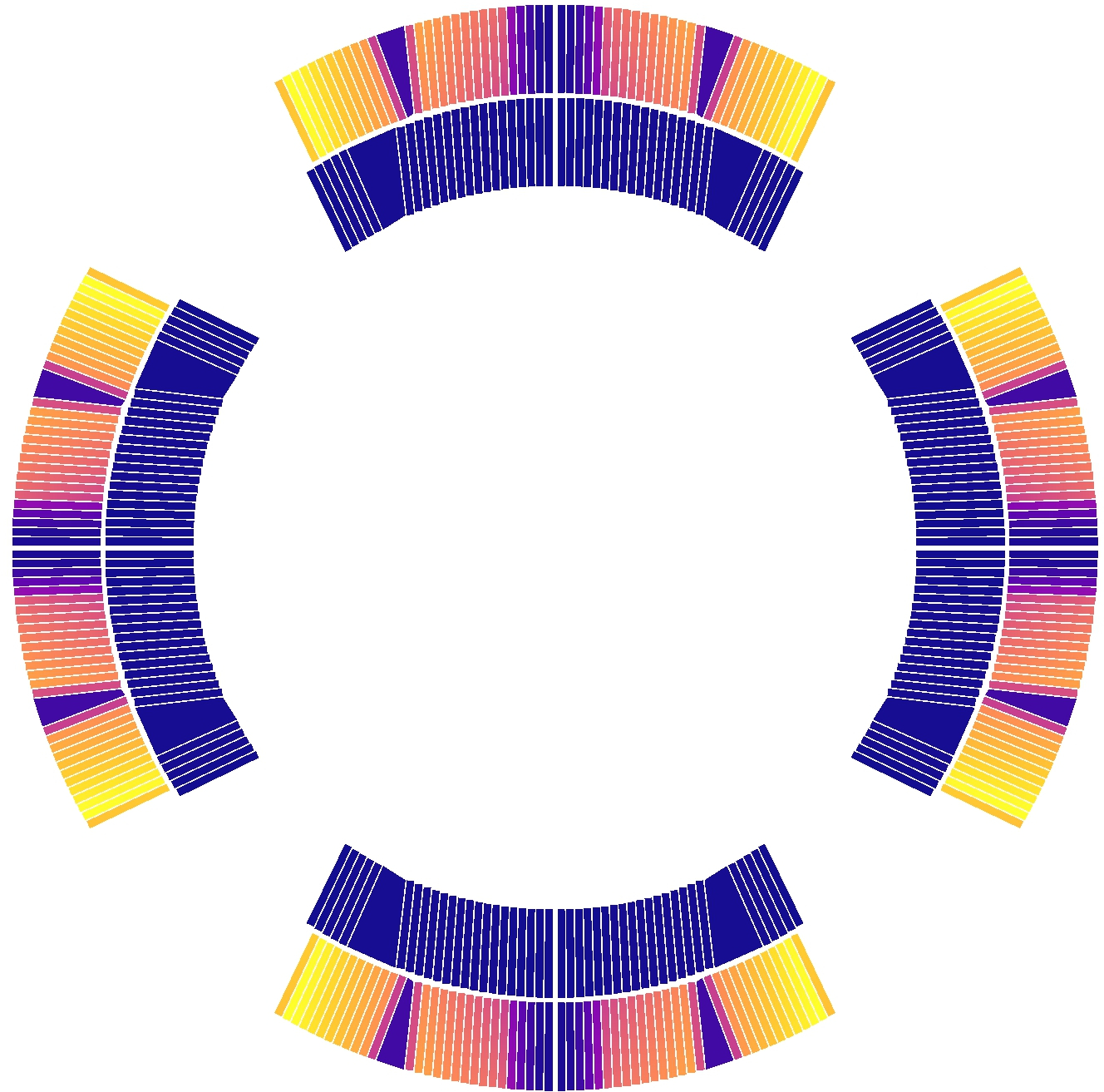};
    \end{axis}
\end{tikzpicture}%
        \caption{MQXF temperature.}
        \label{fig:mqxf_th}
    \end{subfigure}
    \hspace*{-0.2cm}
    \begin{subfigure}{.24\linewidth}
        \centering
        \begin{tikzpicture}
    \begin{axis}[
        width=1.45\textwidth,
        axis equal image,
        hide axis,
        enlargelimits=false,
        point meta min = 0,
        point meta max = 12.4,
        colormap name=viridis,
        colorbar horizontal,
        colorbar style={
            title=Magnetic Flux Density $\left(\si{\tesla}\right)$,
            title style={
                at={(0.5,-2)},
                anchor=north,
                font=\footnotesize
            },
            scaled x ticks=false,
            xtick style={draw=none},
            xticklabel style = {font=\footnotesize},
            xtick={0, 12.4},
            at={(0.5,-0.025)},
            anchor=north,
            width = .5\textwidth,
        },
        colorbar/width=0.25cm,
    ] 
        \addplot graphics[xmin=0,xmax=1301,ymin=0,ymax=1304] {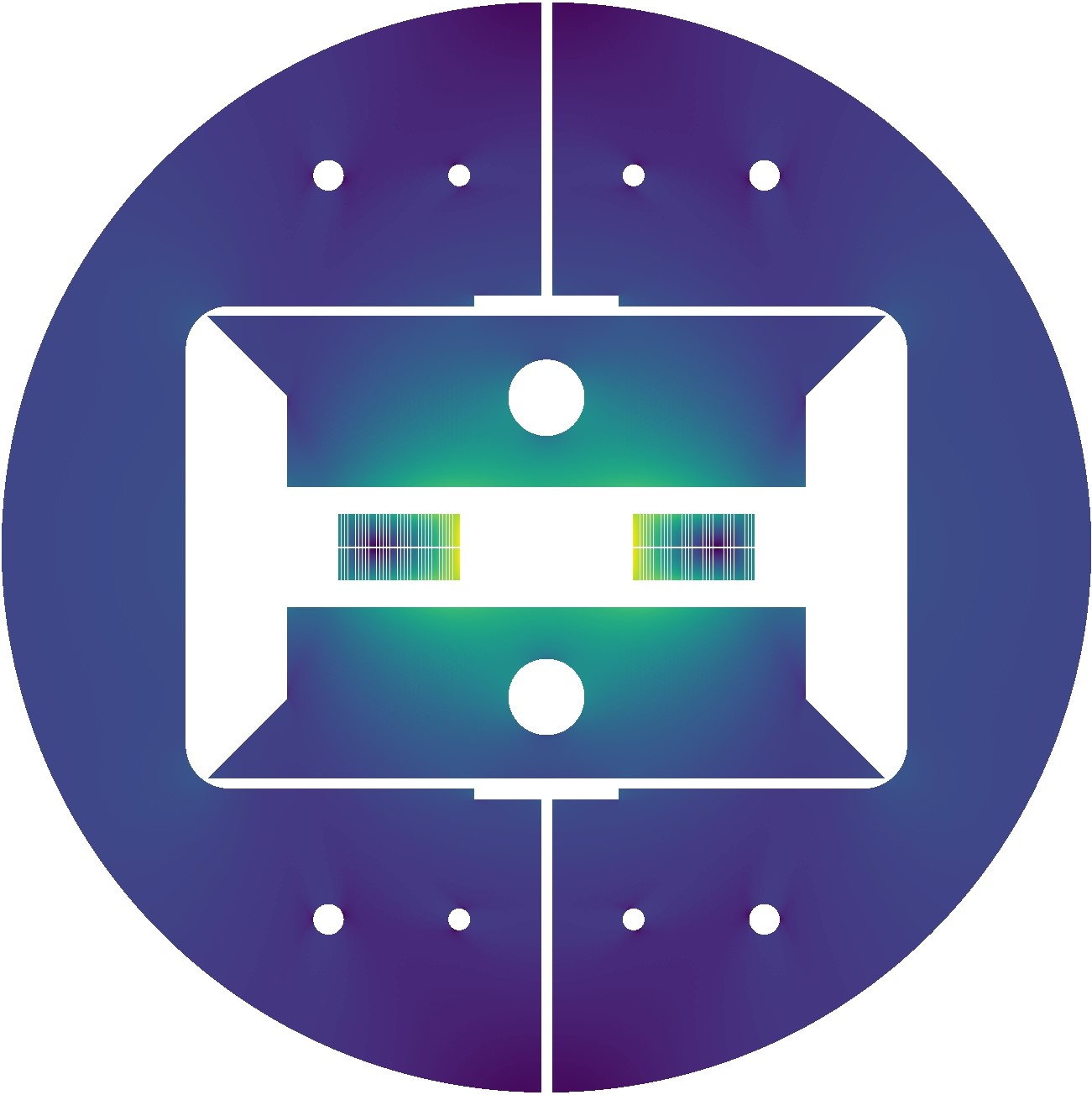};
    \end{axis}
\end{tikzpicture}%
        \caption{SMC flux density.}
        \label{fig:smc_em}
    \end{subfigure}
    \hspace*{0.1cm}
    \begin{subfigure}{.24\linewidth}
        \centering
        \begin{tikzpicture}
    \begin{axis}[
        width=1.45\textwidth,
        axis equal image,
        hide axis,
        enlargelimits=false,
        point meta min = 1.9,
        point meta max = 303,
        colormap/plasma,
        colorbar horizontal,
        colorbar style={
            title=Temperature $\left(\si{\kelvin}\right)$,
            title style={
                at={(0.5,-2)},
                anchor=north,
                font=\footnotesize
            },
            scaled x ticks=false,
            xtick style={draw=none},
            xticklabel style = {font=\footnotesize},
            xtick={1.9, 303},
            at={(0.5,-2.7)},
            anchor=north,
            width = .5\textwidth,
        },
        colorbar/width=0.25cm,
    ] 
        \addplot graphics[xmin=0,xmax=2337,ymin=0,ymax=378] {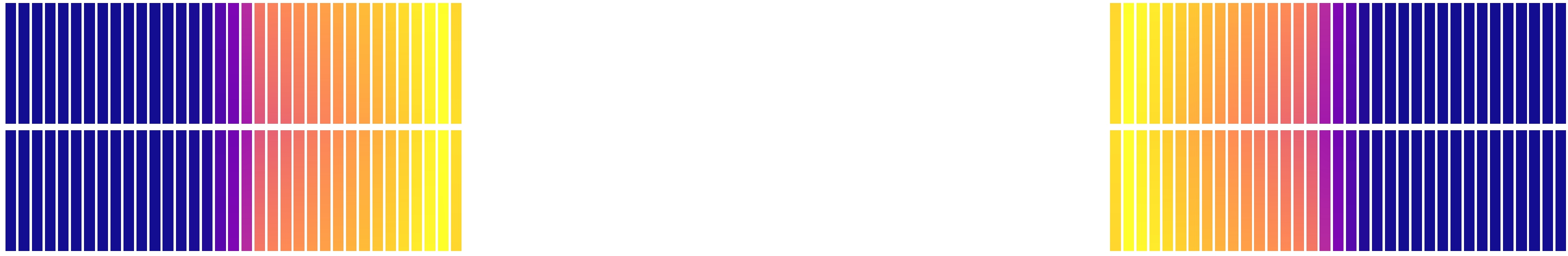};
    \end{axis}
\end{tikzpicture}%
        \caption{SMC temperature.}
        \label{fig:smc_th}
    \end{subfigure}
    \caption{Magnetic flux density of the magnetostatic solution followed by thermal transient solution for two LHC cos-theta magnets with Nb-Ti cables: low-beta quadrupole MQXA \cite{AJIMA2005499, Ogitsu_2002aa, Ostojic_2002aa, Yamamoto_1997aa, LHCdesign} (\ref{fig:mqxa_th}, \ref{fig:mqxa_em}); double-aperture main bending dipole MB \cite{Rossi2003, Rossi_2003aa, Rossi_2004aa, Fessia_2000aa, LHCdesign} (\ref{fig:mb_th}, \ref{fig:mb_em}), and two magnets with Nb$_3$Sn cables: HL-LHC low-beta quadrupole MQXF \cite{Ferracin_2010aa, Ferracin_2014aa, Ambrosio_2015aa, Izquierdo-Bermudez_2017aa, Ravaioli_2018aa, Ferracin2019, Izquierdo-Bermudez_2021aa} (\ref{fig:mqxf_th}, \ref{fig:mqxf_em}); block-coil dipole SMC \cite{Bajko2012, Regis_2010aa, Perez_2015aa} (\ref{fig:smc_th}, \ref{fig:smc_em}). For the thermal transient solution, QHs are considered and the simulation is conducted until one half turn reaches \SI{300}{\kelvin}, the last time step is shown for the thermal simulation. The thermal transient results are obtained for a coupling with a magnetostatic solution for time-invariant operating currents.
    }
    \label{fig:magnets_1}
\end{figure*}

The TSA algorithm was developed with \texttt{FiQuS}'s emphasis on automation and parametrisation in mind. As a result, thermal transient TSA models are available for all multipole magnets supported in \texttt{FiQuS}. As a fourth case study, \Fref{fig:magnets_1} presents results of magnetostatic-thermal transient simulations of four magnets in order to showcase a range of the capabilities of the \texttt{FiQuS} multipole module. Results are shown for two LHC cos-theta magnets based on Nb-Ti conductors, an HL-LHC cos-theta magnet, and a block-coil based on Nb$_3$Sn conductors. 

All magnets are powered to nominal current and some of their half turns are quenched by powering QHs at $t = \SI{0}{\second}$. All magnet and QH parameters are found in \cite{multipole-analysis}.  The thermal transient simulation is performed with time-invariant operating current in the magnets until $T_\text{hotspot} > \SI{300}{\kelvin}$. The final temperature distributions result from the QH power, the Joule heating in half turns quenched by the magnetic field and temperature distribution, and thermal diffusion in the half turns and TSLs.

For comprehensive quench detection and protection simulations, future work will consider coupling the thermal transient model to a magnetodynamic model. The latter will incorporate an efficient reduced order model that describes the magnetisation and instantaneous power loss in composite superconductors based on \cite{dular2024reducedorderhystereticmagnetization, Dular_2024aa}.

\section{Conclusions}\label{sec:conclusions}

This study demonstrated the effectiveness of the thermal thin-shell approximation (TSA) implemented within \texttt{FiQuS} for simulating thermal transients in superconducting magnets. The two-dimensional TSA method collapses thin insulation layer surfaces into lines, significantly reducing mesh complexity and computational time while maintaining accuracy. The TSA implementation in the \texttt{FiQuS} multipole module was verified for a simplified four-conductor model and a full cross-section of the superconducting dipole magnet MBH by comparison with models with surface mesh insulation. For the MBH magnet, the TSA model achieved results with error in the hotpot temperature of less than \SI{0.5}{\percent} with a speed increase of more than 38 times compared to the surface meshed models. The analysis also highlighted the TSA model's ability to handle non-linear material properties, complex multi-layer configurations, and cryogenic cooling, proving its suitability for quench protection studies. The TSA method was also employed cooperatively with magnetostatic simulations for multi-physics solutions, including quench heaters, as multilayer structures with an internal power source, accurately capturing the quench heater delay compared to measured data. The flexibility of the implementation in the \texttt{FiQuS} multipole module was highlighted by showing coupled magnetostatic and thermal transient solutions of several LHC and HL-LHC magnets.

\section*{Acknowledgment}
The work of E. Schnaubelt is supported by the Graduate School CE within the Centre for Computational Engineering at TU Darmstadt and by the Wolfgang Gentner Programme of the German Federal Ministry of Education and Research (grant no. 13E18CHA).

\printbibliography

\end{document}